\documentclass[onecolumn,11pt]{article}
\usepackage{mathrsfs}
\usepackage{amsmath, alltt}
\usepackage{rotating}
\usepackage{graphicx}
\usepackage{subfigure}
\usepackage[onehalfspacing]{setspace}
\usepackage[hmargin={1.0in, 1.0in}, vmargin={1.0in, 1.0in}]{geometry}
\usepackage{amsfonts}
\usepackage{amssymb}
\usepackage{natbib}
\usepackage{amsmath}
\usepackage{bm}
\usepackage{latexsym}
\usepackage{amssymb,bm}
\usepackage[section]{placeins}
\usepackage{longtable}
\usepackage{lscape}
\usepackage{scalefnt}
\usepackage{setspace}
\usepackage{color}
\usepackage{hyperref}[]
\usepackage{enumerate}
\hypersetup{
	colorlinks=true,
	linkcolor=blue,
	filecolor=magenta,
	urlcolor=cyan,
	citecolor=blue,
}
\usepackage{cleveref}
\usepackage{xcolor}
\usepackage{algorithm}
\usepackage{algpseudocode}
\usepackage[title]{appendix}

\usepackage{titlesec}
\titlespacing*\section{0pt}{8pt plus 2pt minus 2pt}{0pt plus 2pt minus 2pt}
\titlespacing*\subsection{0pt}{6pt plus 2pt minus 2pt}{0pt plus 2pt minus 2pt}
\titlespacing*\subsubsection{0pt}{2pt plus 2pt minus 2pt}{0pt plus 2pt minus 2pt}

\usepackage{xr}
\def\bfbeta{\boldsymbol{\beta}}
\def\bfrho{\boldsymbol{\rho}}
\def\bfzeta{\boldsymbol{\zeta}}
\def\bftheta{\boldsymbol{\theta}}

\DeclareMathOperator*{\argmax}{arg\,max}
\DeclareMathOperator*{\argmin}{arg\,min}

\begin{document}
\setlength{\abovedisplayskip}{3pt}
\setlength{\belowdisplayskip}{3pt}
\setlength{\abovedisplayshortskip}{1pt}
\setlength{\belowdisplayshortskip}{1pt}

%
\title{Enhanced Pricing and Management of Bundled Insurance Risks with Dependence-aware Prediction using Pair Copula Construction}

\author{
    Peng Shi \\
    Wisconsin School of Business\\
    University of Wisconsin-Madison\\
    \and
    Zifeng Zhao \\
    Mendoza College of Business\\
    University of Notre Dame\\
}
\date{}
\maketitle
\vspace{-1.5cm}
\begin{abstract}
We propose a dependence-aware predictive modeling framework for multivariate risks stemmed from an insurance contract with bundling features -- an important type of policy increasingly offered by major insurance companies. The bundling feature naturally leads to longitudinal measurements of multiple insurance risks, and correct pricing and management of such risks is of fundamental interest to financial stability of the macroeconomy. We build a novel predictive model that fully captures the dependence among the multivariate repeated risk measurements. Specifically, the longitudinal measurement of each individual risk is first modeled using pair copula construction with a D-vine structure, and the multiple D-vines are then integrated by a flexible copula. While our analysis mainly focuses on multivariate insurance risks, the proposed model indeed contributes to the broad research area of longitudinal data analysis. In particular, it provides a unified modeling framework for multivariate longitudinal data that can accommodate different scales of measurements, including continuous, discrete, and mixed observations, and thus can be potentially useful for various economic studies. A computationally efficient sequential method is proposed for model estimation and inference, and its performance is investigated both theoretically and via simulation studies. In the application, we examine multivariate bundled risks in multi-peril property insurance using proprietary data from a commercial property insurance provider. The proposed model is found to provide improved decision making for several key insurance operations. For underwriting, we show that the experience rate priced by the proposed model leads to a 9\% lift in the insurer's net revenue. For reinsurance, we show that the insurer underestimates the risk of the retained insurance portfolio by 10\% when ignoring the dependence among bundled insurance risks.
\end{abstract}

\noindent{\bf Keywords:}  Multivariate longitudinal data, Copula, D-Vine, Insurance operations, Predictive analytics, Graphical model 

\newpage

\section{Introduction}
In the past decade, insurance industry, especially property and casualty insurance, has been advancing the use of predictive analytics to leverage big data and improve business performance. Statistical learning of insurance risks has become an essential component in the data-driven decision making in various insurance operations~\citep{Frees2015}. This study focuses on the predictive modeling for nonlife insurance products with a \textit{bundling} feature. 

Bundling is an increasingly popular design in modern short-term insurance contracts and can take different forms in practice: a comprehensive auto insurance policy provides coverage for both collision and third-party liability; employee compensation insurance provides benefits for wage replacement, medical treatment, and vocational rehabilitation; an open peril property insurance policy covers losses due to all types of causes subject to certain exclusions. As another prominent example, in personal lines of business, auto insurance and homeowner insurance are often marketed to households as a package. The bundling feature naturally leads to longitudinal measurements of multivariate risks, where the insurer observes multiple risk outcomes of an insurance contract over time. This serves as motivation of our study and we aim to propose a general framework of predictive modeling for multivariate insurance risks with longitudinal/repeated measurements.

An essential element of predictive modeling is to accurately assess the dependence/association among insurance risks, which helps track the evolution and thus generate prediction of future risks. Two types of dependence are of primary interests to insurers and have been studied in separate strands of literature. The first is the temporal dependence of a single insurance risk. The availability of longitudinal measurements allows an insurer to adjust a policyholder's premium based on the claim history, known as experience rating in insurance~\citep{Pinquet2013}. The experience rate is determined by the insurer's ability of learning hidden risks of policyholders that evolve over time (see e.g.\ \cite{frees2006copula}, \cite{BoucherInoussa2014} and \cite{oh2020bonus}). The second is the contemporaneous dependence among multiple insurance risks. Correlated claims induces concentration risk in the insurance portfolio on the liability side of an insurer's book, 
thus understanding the effects of dependence in risk aggregation can provide valuable guidance to the insurer's operation in claim management (\cite{FreesValdez2008} and \cite{jessup2020fitting}) and capital management (\cite{bernard2014risk} and \cite{wang2019dual}).

The multivariate longitudinal measurements of insurance risks from bundling contracts naturally involve both temporal and contemporaneous dependence. Thus, a predictive model that allows a joint analysis of the two types of dependence can deliver unique insights to an insurer's operation and enable an insurer to perform prospective experience rating, determine optimal risk retention, and estimate the amount of risk capital in a coherent and consistent manner. However, despite the appealing benefits, the majority of existing models in the literature are only capable of examining the two types of dependence separately. The scarcity of a unified and flexible predictive modeling framework for multivariate insurance risks is due to several challenges.


A notable challenge is the discreteness in the risk measurements. A common measurement of a policyholder's risk is the annual number of claims incurred~\citep{denuit2007actuarial}. Thus, in the context of bundling insurance products, we need to jointly model multivariate claim counts stemmed from each of the multiple insurance risks covered in the contract. Due to the discreteness of claim count, the classical concept of correlation and existing modeling and analysis techniques for multivariate continuous data are not applicable~\citep{Joe2014}. Another challenge in predictive modeling is to determine the optimal number of historical observations to be used in the prediction for future risks. Existing literature mostly models temporal dependence among insurance outcomes by borrowing techniques (notably the autoregressive models) from the time series forecasting literature. However, predictive models and specification tests of optimal order for time series data~(which consists of hundreds of observations along time dimension) are generally not appropriate for longitudinal data~(which only contains a handful of temporal observations).

We remark that several strategies have been studied for modeling multivariate longitudinal data in the biostatistics literature, see \cite{Verbeke2014} and \cite{farewell2017two} for recent reviews. However, these methods are typically built upon random effect or latent variable models, which offer limited modeling choices for marginal behavior and only allows for specific structures on the temporal-contemporaneous dependence. Moreover, these methods typically are designed for continuous data and become computationally infeasible for insurance claim counts, and discrete data in general, especially in the setting of nonstandard marginal distributions. To conserve space, we refer to Section \ref{subsec:add_litreview} of the supplementary material for more detailed review of this literature.

To fill the gap in the literature, we propose a unified predictive modeling framework for multivariate longitudinal measurements of insurance risks that simultaneously accommodates the temporal and contemporaneous dependence. Specifically, we integrate a generalized linear regression based framework with a flexible graphical model named vine~\citep{BedfordCooke2002, Aas2009}, where the longitudinal measurement of each individual risk is first modeled using pair copula construction with a D-vine structure, and the multiple D-vines are then linked together via a flexible multivariate copula. We refer to Section \ref{sec:method} for the technical definition and more detailed literature review of copula and vine. 


We remark that as a fundamental tool for modeling dependence, copula has been widely studied in the econometrics literature, see \cite{Chen2006a, Chen2006, Chen2009, Beare2010, Oh2013, Oh2017, Chen2021} for representative works. However, all these works primarily focus on multivariate time series modeling (with continuous observations) and cannot be easily adapted to model multivariate longitudinal data. Indeed, to our best knowledge, copula-based models for multivariate longitudinal data are still scarce in the literature.

Our work provides one of the first effort in constructing copula-based models for multivariate longitudinal data. Due to the use of copula, the proposed framework allows for separate specification of the marginal regression model from the dependence model, and thus enables unified accommodation for different scales of multivariate longitudinal data, including continuous, discrete, and mixed outcomes. 
Moreover, the D-vine based pair copula construction is flexible and further allows the proposed model to specify the temporal and contemporaneous dependence unrestrictedly, and thus can achieve a wide range and various types of dependence among multivariate longitudinal measurements. Thanks to these important properties, the proposed model can be useful for modeling multivariate longitudinal data commonly encountered under various economic studies and thus contributes to the broad research area of longitudinal data analysis. We further develop diagnostic checks for model specification and propose a novel data-driven procedure that automatically determines the optimal weight given to historical observations for future risk prediction. We propose a computationally efficient sequential method for the estimation and inference of the proposed model and investigate its performance both theoretically and via simulation studies. 

Compared to standard modeling practice in the insurance industry, the proposed predictive model achieves simultaneous modeling of both temporal and contemporaneous dependence among bundled insurance risks and provides dependence-aware prediction that is shown to bring significant values to key insurance operations such as risk segmentation and risk management. Specifically, using data from a Wisconsin property insurance provider, we show that the risk pricing derived from the proposed predictive model provides a 9\% lift of the insurer's profit in the underwriting and ratemaking operation, and the proposed model provides more truthful risk assessment of the retained insurance portfolio of the insurer by 10\% in the reinsurance operation.

The rest of the paper is structured as follows. Section \ref{sec:generalsetting} highlights the essential role of predictive models in improving decision making for insurance operations. Section \ref{sec:method} introduces the pair copula construction based unified modeling framework for multivariate longitudinal data. Section \ref{sec:inference} proposes a sequential inference procedure for model selection and estimation, and further establishes its theoretical guarantees and performs simulation studies. Section \ref{sec:empirical} conducts empirical analysis of multivariate claim counts from a large-scale Wisconsin property insurance program. Section \ref{sec:manage} illustrates managerial implications and significance of the proposed dependence-aware predictive model in key insurance operations. Section \ref{sec:conclusion} concludes. Technical materials, additional simulation experiments and data analysis results are gathered in the supplementary material.

\section{Predictive Modeling in Key Insurance Operations}\label{sec:generalsetting}

In this section, we discuss the use of predictive modeling in insurance operations and the decision-makings faced by insurers for managing a portfolio of insurance contracts with bundled risks.

Denote $y_{it}^{(j)}$ as the measurement~(such as annual claim count) of the $j$th ($=1,\cdots,J$) insurance risk for policyholder $i$ ($=1,\ldots,n$) in period $t$ ($=1,\ldots,T$). The number of observations $T$ is typically small, partly due to the short-term nature of nonlife insurance contracts. Let $\bm{x}_{it}^{(j)}$ be the set of exogenous predictors the insurer uses for evaluating the $j$th risk of policyholder $i$ in period $t$, such as characteristics of the policyholder or the contract~(see Table \ref{tab:covariates} in the supplementary material for an example). Define $\bm{x}_{i}^{(j)}=\{\bm{x}_{it}^{(j)}: t\geq 1\}$. Define $H_{it}^{(j)} = \bm{x}_{i}^{(j)}\bigcup\{y_{is}^{(j)}: s=1,\ldots,t\}$, which includes the history of the $j$th risk up to period $t$ for policyholder $i$, together with the exogenous predictors. Note that since the predictors are exogenous, without loss of generality, we can assume that we have the knowledge of $\bm{x}_{i}^{(j)}$ at all time periods $t$ for notational simplicity.

Statistically speaking, we define the predictive model for the multivariate insurance risks $\bm{y}_{i,T+1}=(y^{(1)}_{i,T+1},\ldots,y^{(J)}_{i,T+1})$ of policyholder $i$~($=1,\cdots,n$) in the future period $T+1$ as
\begin{align} \label{equ:predm}
F\left(\bm{y}_{i,T+1}|H_{iT}^{(1)},\ldots,H_{iT}^{(J)}\right)=F\left(y^{(1)}_{i,T+1},\ldots,y^{(J)}_{i,T+1}|H_{iT}^{(1)},\ldots,H_{iT}^{(J)}\right),
\end{align}
which is the conditional joint distribution of future multivariate insurance risks $\bm{y}_{i,T+1}$ given its claim history and exogenous predictors. Importantly, the conditional distribution \eqref{equ:predm} involves both temporal and contemporaneous dependence among $\{\bm{y}_{i,1},\bm{y}_{i,2},\cdots,\bm{y}_{i,T+1}\}$. We show below that an accurate modeling of \eqref{equ:predm} is essential for informed decision making in key insurance operations. 


Assuming the measurement of risk $y_{it}^{(j)}$ represents the claim count, we can express the aggregate losses of the bundled insurance contract stemmed from policyholder $i$ in period $T+1$ as:
\begin{align} \label{equ:indagg}
S_{i,T+1}=\alpha_{i,T+1}^{(1)}y_{i,T+1}^{(1)}+\cdots+\alpha_{i,T+1}^{(J)}y_{i,T+1}^{(J)},
\end{align}
where $\alpha_{i,T+1}^{(j)}$ represents the average cost per claim for the $j$th risk of policyholder $i$. 
In this study, our primary focus is the dependence among claim frequency, we thus assume one has prior knowledge on the average claim cost~\citep{DeJongHeller2008}.

The first insurance operation of interest is regarding risk segmentation, specifically renewal underwriting and ratemaking. Since nonlife insurance contracts in general have short duration, the insurer has the opportunity to decide whether to provide coverage~(underwriting) when the current contract expires and is subject to renewal. Provided that the insurer accepts the risk, a new premium is to be determined~(ratemaking). Ratemaking at renewal is based on two types of information, the exogenous rating variables and the policyholder's claim history. Adjusting premiums using the claim history is known as experience rating. In the actuarial statistics literature, credibility theory has long been introduced as a comprehensive way of incorporating loss experience into pricing (see \cite{buhlmann2006course}). \cite{frees1999longitudinal} established the connection between the credibility theory and longitudinal data models.

Given a deductible $d\geq 0$, a risk retention feature commonly found in property and health insurance contracts, the experience rate is based on the expected loss cost in period $T+1$
\begin{align} \label{equ:exprate}
\mathbb{E}\left((S_{i,T+1}-d)_+|H_{iT}^{(1)},\ldots,H_{iT}^{(J)}\right),
\end{align}
where $(S_{i,T+1}-d)_{+}=\max\{S_{i,T+1}-d,0\}$. The insurer uses the experience rate \eqref{equ:exprate} to decide whether to provide coverage and to adjust the premium upon acceptance of the risk. The expected loss in \eqref{equ:exprate} depends on $S_{i,T+1}$, whose behavior is effectively determined by the conditional distribution \eqref{equ:predm}, manifesting the crucial value of accurate predictive modeling of temporal and contemporaneous dependence among bundled insurance risks.

The second insurance operation under consideration is reinsurance. As a risk management tool, insurers use reinsurance to reduce and stabilize the cost of insurance while taking into account its own risk transfer capacity. In particular, the insurer often cedes some of its portfolio risk to an reinsurer to reduce its liability when it is subject to underwriting capacity~\citep{albrecher2017reinsurance}. We examine the quota share treaty, a form of pro-rata reinsurance popular in the insurance industry, in which the reinsurer assumes an agreed percentage of each insurance risk in the portfolio and shares all premiums and losses accordingly with the insurer.

For an insurer's portfolio of $n$ bundled insurance contracts $\{S_{i,T+1}:i=1,\ldots,n\}$, we consider the general case of quota share treaty where the retention can vary by contract. Let $\delta_i$ denote the retained quota for the $i$th contract, the retained risk of the insurer can be represented by
\begin{align} \label{equ:rein}
S_{T+1}^{*} = \delta_1 S_{1,T+1} + \cdots + \delta_n S_{n,T+1},
\end{align}
where $S_{i,T+1}$ is defined in (\ref{equ:indagg}). The insurer's goal is to find optimal amount of retention $\{\delta_1,\cdots,\delta_n\}$ across the $n$ contracts to minimize volatility of the retained risk $S_{T+1}^*$ under constraints on its underwriting capacity $\mathbb{E}(S_{T+1}^*)$. It is evident that both temporal and contemporaneous dependence among the multivariate insurance risks in \eqref{equ:predm} directly impact the behavior of $S_{i,T+1}$ and thus are critical to the calculation of optimal risk retention $\delta_i$ in the portfolio risk management.

\section{Methodology}\label{sec:method}
To model longitudinal measurements of multivariate insurance risks, we propose an approach that integrates a regression-based framework with a flexible graphical model named vine~\citep{BedfordCooke2001,BedfordCooke2002}, also known as pair copula construction~\citep{Aas2009} in the literature.

The vine model can be seen as a special type of copula. A $d$-dimensional copula is a multivariate distribution function on $(0, 1)^d$ with uniform margins. By the celebrated \cite{Sklar1959}'s theorem, any multivariate distribution $F$ can be separated into its marginals $(F_1,\ldots, F_d)$ and a copula $C$, where the copula captures all the scale-free dependence of the multivariate distribution. In particular, denote $\mathbf Z \in \mathbb{R}^d$ as a random vector following a multivariate distribution $F$, we have $F(z_1,\ldots, z_d)=C(F_1(z_1), \ldots, F_d(z_d))$ for any $(z_1,\ldots, z_d) \in \mathbb{R}^d$.

The key feature of vine is that it provides a systematic framework of building flexible multivariate distributions based on a sequence of \textit{bivariate} copulas. The original literature of vine mainly focused on modeling multivariate data with continuous outcomes, see \cite{KurowickaCooke2006} and \cite{Aas2009} for early works. Vine has received extensive attention in the recent literature of dependence modeling, due to its flexibility especially in terms of modeling tail dependence and asymmetric dependence~(see e.g.\ \cite{JoeKurowicka2011}). Building upon the framework for continuous outcomes, \cite{PanagiotelisCzadoJoe2012} introduced pair copula construction for discrete data, \cite{Stober2015} examined the vine model for multivariate responses with both continuous and discrete variables, \cite{ShiYang2018} employed pair copula construction to model temporal dependence among univariate longitudinal data with mixed outcomes, and \cite{Barthel2018} further extended the vine approach to event time data with censoring. However, all these studies focus on univariate longitudinal data or multivariate observations without repetition, pair copula construction for multivariate longitudinal data with multilevel structure is sparse in the literature. Recently, \cite{BrechmannCzado2015} and \cite{Smith2015} discussed possible strategies of constructing vines for multivariate time series. However, the models therein only accommodates continuous observations and cannot be easily adapted to model multivariate longitudinal data. 

Based on pair copula construction, we propose a unified modeling framework for multivariate longitudinal data in Section \ref{subsec:vine_framework} and further present its detailed application for modeling and predicting longitudinal observations of multivariate insurance claim counts in Sections \ref{subsec:riskmodel} and \ref{subsec:prediction}.

\subsection{A Unified Modeling Framework}\label{subsec:vine_framework}
In this section, we present a unified modeling framework for multivariate longitudinal data that can accommodate different scales of measurements, including continuous, discrete, and mixed outcomes.

Following notations in Section \ref{sec:generalsetting}, for policyholder $i=1,\cdots,n$, we denote $\bm{H}_{it}=(H_{it}^{(1)},\ldots,H_{it}^{(J)})$ as its claim history up to time $t$, denote $\bm{X}_i=\{\bm{x}_i^{(1)},\cdots,\bm{x}_i^{(J)}\}$ as the collection of its exogenous predictors, and further define $\bm{H}_{i0}=\bm{X}_i.$ In the following, we use $f$ to denote the (conditional) pmf or pdf for either univariate or multivariate distributions, and use $F$ to denote the corresponding cdf. Denote $\bm{y}_{it}=(y_{it}^{(1)},\cdots,y_{it}^{(J)})$. Given the exogenous predictors $\bm{X}_i$, the joint distribution of the observed multivariate insurance risks $\{\bm{y}_{i1},\ldots,\bm{y}_{iT}\}$ for policyholder $i$ can be written as
\begin{align}  \label{equ:joint}
f(\bm{y}_{i1},\ldots,\bm{y}_{iT}|\bm{X}_i) =f(\bm{y}_{i1}|\bm{X}_i)f(\bm{y}_{i2}|\bm{y}_{i1};\bm{X}_i)\cdots f(\bm{y}_{iT}|\bm{y}_{i1},\ldots,\bm{y}_{iT-1};\bm{X}_i) =\prod_{t=1}^{T}f(\bm{y}_{it}|\bm{H}_{i,t-1}).
\end{align}
Note that the conditional decomposition in \eqref{equ:joint} is generic and does not impose any constraint on the model specification of $(\bm{y}_{i1},\ldots,\bm{y}_{iT})|\bm{X}_i$. In other words, the decomposition \eqref{equ:joint} applies to all scales of measurements, including continuous, discrete, and mixed outcomes.

By the \cite{Sklar1959}'s theorem and its extension~\citep{Patton2006}, there exists a $J$-variate copula $C^{J}$ such that the corresponding cdf of the joint distribution of $\bm{y}_{it}=(y_{it}^{(1)},\ldots,y_{it}^{(J)})$ conditioning on $\bm{H}_{i,t-1}$ in (\ref{equ:joint}) can be further represented as
\begin{align} \label{equ:cjoint}
F(\bm{y}_{it}|\bm{H}_{i,t-1}) = C^{J}\left(F\left(y_{it}^{(1)}|H_{i,t-1}^{(1)}\right),\ldots,F\left(y_{it}^{(J)}|H_{i,t-1}^{(J)}\right)\right),
\end{align}
for $t=1,\cdots,T$. The copula $C^{J}$ captures the contemporaneous dependence among the multivariate insurance risks $\bm{y}_{it}=(y_{it}^{(1)},\ldots,y_{it}^{(J)})$, and the conditional marginal distribution $F(y_{it}^{(j)}|H_{i,t-1}^{(j)})$ captures the temporal dependence within the $j$th individual insurance risk and characterizes the behavior of $y_{it}^{(j)}$ conditioned on its claim history and exogenous predictors, for $j=1,\cdots,J$. It is important to point out that, according to the definition in \eqref{equ:predm}, we can construct a predictive model for bundled insurance risks based on \eqref{equ:cjoint} by setting $t=T+1$.

\begin{figure}[htp]
  \begin{center}
   \includegraphics[width=0.8\textwidth,angle=0]{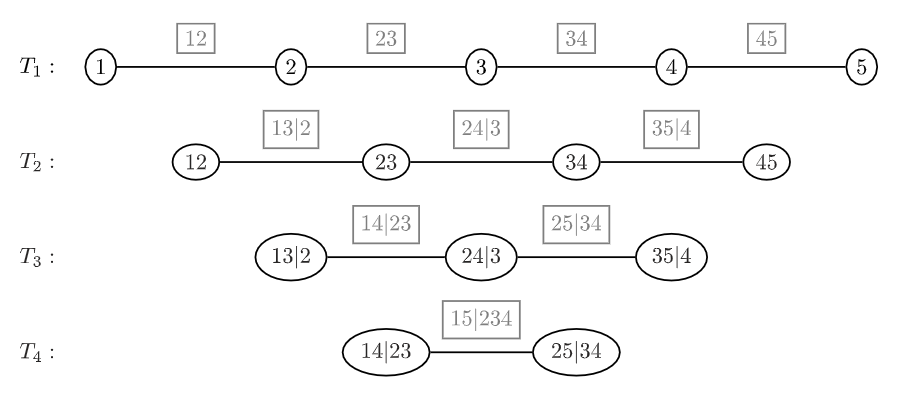}
  \end{center}
  \vspace{-0.8cm}
   \caption{Graphical representation of a fully specified 5-dimensional D-vine.}
   \label{fig:dvine}
\end{figure}

Note that the conditional distribution $F(y_{it}^{(j)}|H_{i,t-1}^{(j)})$ in (\ref{equ:cjoint}) can be derived from the joint distribution of $(y_{i1}^{(j)},\ldots,y_{iT}^{(j)})\big|\bm{x}_i^{(j)}$ for $j=1,\ldots,J$. For flexibility of the predictive model, we construct this joint distribution via pair copula construction based on a D-vine structure. Figure \ref{fig:dvine} shows a graphical representation of a D-vine for $T=5$. The fully specified D-vine contains $T-1=4$ trees $T_1,\cdots,T_4$. In each tree, only adjacent nodes are connected by an edge, and edges become nodes in the next tree. Each node represents a (conditional) distribution and the edge indicates a bivariate copula linking the distributions of the two connecting nodes. The edges of the entire D-vine summarize the bivariate copulas that contribute to the pair copula construction. We later provide more concrete and detailed discussion of D-vine.

Based on the graphical representation of D-vine, the joint distribution of $(y_{i1}^{(j)},\ldots,y_{iT}^{(j)})\big|\bm{x}_i^{(j)}$ can be nicely written as a factor form such that
\begin{align} \label{equ:jtime}
&f\left(y_{i1}^{(j)},\ldots,y_{iT}^{(j)}|\bm{x}_i^{(j)}\right) \nonumber\\
=&\prod_{t=1}^{T}f\left(y_{it}^{(j)}|\bm{x}_i^{(j)}\right)\prod_{t=2}^{T}\prod_{s=1}^{t-1}\frac{f\left(y^{(j)}_{is},y^{(j)}_{it}|y^{(j)}_{is+1},\ldots,y^{(j)}_{it-1};\bm{x}_i^{(j)}\right)}{f\left(y^{(j)}_{is}|y^{(j)}_{is+1},\ldots,y^{(j)}_{it-1};\bm{x}_i^{(j)}\right)f\left(y^{(j)}_{it}|y^{(j)}_{is+1},\ldots,y^{(j)}_{it-1};\bm{x}_i^{(j)}\right)}.
\end{align}
Here and after, for two indices $e_1,e_2 \in\{1,2,\cdots,T\}$, we use the convention that $(y^{(j)}_{ie_1},\ldots,y^{(j)}_{ie_2})=\varnothing$ if $e_2>e_1.$ We remark that, same as \eqref{equ:joint}, the decomposition in \eqref{equ:jtime} is generic and does not impose any constraint on the joint distribution of $(y_{i1}^{(j)},\ldots,y_{iT}^{(j)})\big|\bm{x}_i^{(j)}$, and thus accommodates all scales of measurements, including continuous, discrete, and mixed outcomes.

For more intuition, we map the decomposition in \eqref{equ:jtime} to Figure \ref{fig:dvine}. Specifically, the nodes in tree 1~(i.e.\ $T_1$) represent the univariate marginal distribution $f(y_{it}^{(j)}|\bm{x}_i^{(j)})$ of $y_{it}^{(j)}$ observed at each time $t=1,\cdots,T$. For $t=2,\cdots,T$ and $s=1,\cdots,t-1$, from left to right, the $s$th edge in tree $t-s$~(i.e.\ $T_{t-s}$) of Figure \ref{fig:dvine} corresponds to a bivariate copula, denoted by $C^{(j)}_{s, t}$. The bivariate copula $C^{(j)}_{s, t}$ is used to specify the conditional joint distribution $f(y^{(j)}_{is},y^{(j)}_{it}|y^{(j)}_{is+1},\ldots,y^{(j)}_{it-1};\bm{x}_i^{(j)})$ in \eqref{equ:jtime} by linking the $s$th and $(s+1)$th nodes in tree $t-s$, which record the conditional distributions $f(y^{(j)}_{is}|y^{(j)}_{is+1},\ldots,y^{(j)}_{it-1};\bm{x}_i^{(j)})$ and $f(y^{(j)}_{it}|y^{(j)}_{is+1},\ldots,y^{(j)}_{it-1};\bm{x}_i^{(j)})$ respectively. In other words, the $s$th edge in tree $t-s$ of the D-vine represents a bivariate copula $C^{(j)}_{s, t}$, which characterizes the joint behavior of $y^{(j)}_{is}$ and $y^{(j)}_{it}$ conditional on observations $y^{(j)}_{is+1},\ldots,y^{(j)}_{it-1}$ in between.

Thanks to the use of D-vine, the decomposition in \eqref{equ:jtime} separates the specification of the joint distribution $f(y_{i1}^{(j)},\ldots,y_{iT}^{(j)}|\bm{x}_i^{(j)})$ into marginal distributions $\{f(y_{it}^{(j)}|\bm{x}_i^{(j)})\}_{t=1}^T$ and bivariate copulas $\{\{C^{(j)}_{s,t}\}_{s=1}^{t-1}\}_{t=2}^T.$ Moreover, $\{f(y_{it}^{(j)}|\bm{x}_i^{(j)})\}_{t=1}^T$ and $\{\{C^{(j)}_{s,t}\}_{s=1}^{t-1}\}_{t=2}^T$ can fully characterize the joint distribution $f(y_{i1}^{(j)},\ldots,y_{iT}^{(j)}|\bm{x}_i^{(j)})$~(see more details in Section \ref{subsec:riskmodel}). The marginals $\{f(y_{it}^{(j)}|\bm{x}_i^{(j)})\}_{t=1}^T$ can be specified via flexible regression models and the bivariate copulas  $\{\{C^{(j)}_{s,t}\}_{s=1}^{t-1}\}_{t=2}^T$ can be specified to generate a wide range of temporal dependence, as there are various choices for parametric bivariate copulas, such as the elliptical family~(e.g.\ Gaussian and $t$-copula), the Archimedean family~(e.g.\ Clayton, Frank, Gumbel, Joe copula), and the extreme-value copula family, among others. In Section \ref{subsec:riskmodel}, we discuss the detailed specification and computation of each component in \eqref{equ:jtime} for discrete measurements~(i.e.\ annual claim count) of insurance risks.

It is worth mentioning that there are other possible decomposition of $f\left(y_{i1}^{(j)},\ldots,y_{iT}^{(j)}|\bm{x}_i^{(j)}\right)$, such as C-vine and R-vine~\citep{Aas2009}. However, for longitudinal data, the D-vine structure is the most suitable choice, as the observations are arranged in the natural temporal order and the edges of each tree only connect adjacent nodes. These features make D-vine simple to understand and easy to interpret for modeling longitudinal data. 

{\color{black}\textbf{Remark 1}: We note that our D-vine based model for the univariate longitudinal data in \eqref{equ:jtime} is primarily inspired by \cite{ShiYang2018}, which is the first work to employ the method of pair copula construction for modeling temporal dependence in univariate longitudinal insurance claim data. However, as discussed above, our main focus and contribution is the introduction of a copula based unified modeling framework for \textit{multivariate} longitudinal data, where multiple D-vines are further flexibly linked via a cross-sectional copula and thus can accommodate both temporal and contemporaneous dependence. This serves as the key ingredient for modeling and predicting multivariate risks stemmed from an insurance contract with bundling features. In addition, compared to \cite{ShiYang2018}, we formally present and investigate a data-driven approach for the selection of bivariate copulas in the D-vine~(see Section \ref{subsec:inference}) and provide theoretical and numerical studies for the parameter estimation of the proposed D-vine based modeling framework (see Section \ref{subsec:est}).
}


\subsection{Modeling for Multivariate Longitudinal Insurance Claim Counts}  \label{subsec:riskmodel}
In this section, utilizing the unified modeling framework for multivariate longitudinal data in Section \ref{subsec:vine_framework}, we propose a dependence-aware predictive model for multivariate insurance claim counts that accounts for both temporal and contemporaneous dependence among bundled insurance risks. We first specify the D-vine model in \eqref{equ:jtime} for each individual insurance risk $j=1,\cdots, J$ and then specify the multivariate copula $C^J$ that integrates the multiple D-vines in \eqref{equ:cjoint}.

\textbf{The D-vine model}: For the marginal distribution $f(y_{it}^{(j)}|\bm{x}_i^{(j)})$ in \eqref{equ:jtime}, we employ a customized count regression model that can accommodate excess of both zeros and ones, a feature commonly exhibited by insurance claim count data. Specifically, by a slight abuse of notation, we set
\begin{align} \label{equ:mcount}
f\left(y_{it}^{(j)}=y|\bm{x}_i^{(j)}\right)& = {\rm Pr}\left(y_{it}^{(j)}=y|\bm{x}_i^{(j)}\right) =p_{ijt}^{0}\mathbb I(y=0) + p_{ijt}^{1} \mathbb I(y=0) + (1-p_{ijt}^{0}-p_{ijt}^{1})g_{ijt}(y).
\end{align}
We specify $p_{ijt}^{k}$ ($k=0,1$) via a standard multinomial logistic regression such that
\begin{align*}
p_{ijt}^{k} = \frac{\exp\left(\bm{x}_{it}^{(j)\top}\bm{\gamma}_{jk}\right)}{1+ \sum_{k=0}^1\exp\left(\bm{x}_{it}^{(j)\top}\bm{\gamma}_{jk}\right)},~~k=0,1.
\end{align*}
We specify $g_{ijt}(\cdot)$ via a standard negative binomial (NB) regression such that
\begin{align*}
g_{ijt}(y) = \frac{\Gamma(y+\phi_j)}{\Gamma(\phi_j)\Gamma(y+1)}\left(\frac{\mu_{ijt}}{\phi_j+\mu_{ijt}}\right)^{y} \left(\frac{\phi_j}{\phi_j+\mu_{ijt}}\right)^{\phi_j}
\end{align*}
with $\mu_{ijt}=\exp(\bm{x}_{it}^{(j)\top}\bm{\beta}_j)$.

As for the conditional bivariate pmf $f(y^{(j)}_{is},y^{(j)}_{it}|y^{(j)}_{is+1},\ldots,y^{(j)}_{it-1};\bm{x}_i^{(j)})$ in \eqref{equ:jtime}, based on the D-vine structure discussed in Section \ref{subsec:vine_framework}, simple algebra gives
\begin{align} \label{equ:paircopd}
&f\left(y^{(j)}_{is},y^{(j)}_{it}|y^{(j)}_{is+1},\ldots,y^{(j)}_{it-1};\bm{x}_i^{(j)}\right) \nonumber\\
=&C^{(j)}_{s,t}\left(F\left(y^{(j)}_{is}|y^{(j)}_{is+1},\ldots,y^{(j)}_{it-1};\bm{x}_i^{(j)}\right),F\left(y^{(j)}_{it}|y^{(j)}_{is+1},\ldots,y^{(j)}_{it-1};\bm{x}_i^{(j)}\right)\right)\nonumber\\
-&C^{(j)}_{s,t}\left(F\left(y^{(j)}_{is}-1|y^{(j)}_{is+1},\ldots,y^{(j)}_{it-1};\bm{x}_i^{(j)}\right),F\left(y^{(j)}_{it}|y^{(j)}_{is+1},\ldots,y^{(j)}_{it-1};\bm{x}_i^{(j)}\right)\right)\nonumber\\
-&C^{(j)}_{s,t}\left(F\left(y^{(j)}_{is}|y^{(j)}_{is+1},\ldots,y^{(j)}_{it-1};\bm{x}_i^{(j)}\right),F\left(y^{(j)}_{it}-1|y^{(j)}_{is+1},\ldots,y^{(j)}_{it-1};\bm{x}_i^{(j)}\right)\right)\nonumber\\
+&C^{(j)}_{s,t}\left(F\left(y^{(j)}_{is}-1|y^{(j)}_{is+1},\ldots,y^{(j)}_{it-1};\bm{x}_i^{(j)}\right),F\left(y^{(j)}_{it}-1|y^{(j)}_{is+1},\ldots,y^{(j)}_{it-1};\bm{x}_i^{(j)}\right)\right),
\end{align}
where $C^{(j)}_{s,t}$ is the bivariate copula connecting the claim counts $y^{(j)}_{is}$ and $y^{(j)}_{it}$ of risk $j$ in periods $s$ and $t$ (for $s< t$) conditioning on claim counts observed in time periods in between. Graphically speaking, as stated in Section \ref{subsec:vine_framework}, $C^{(j)}_{s,t}$ corresponds to the $s$th edge in tree $t-s$ of the D-vine, linking the $s$th and $(s+1)$th nodes, which record the conditional univariate pmf $f(y^{(j)}_{is}|y^{(j)}_{is+1},\ldots,y^{(j)}_{it-1};\bm{x}_i^{(j)})$ and $f(y^{(j)}_{it}|y^{(j)}_{is+1},\ldots,y^{(j)}_{it-1};\bm{x}_i^{(j)})$ respectively.

As discussed in Section \ref{subsec:vine_framework}, a distinctive advantage of the D-vine is its flexibility to model a wide range of temporal dependence via different choices of the bivariate copulas $\{\{C_{s,t}^{(j)}\}_{s=1}^{t-1}\}_{t=2}^T$. Thus, we do not impose any specific parametric models on $\{\{C_{s,t}^{(j)}\}_{s=1}^{t-1}\}_{t=2}^T$ but instead allow for a data-driven specification of $\{\{C_{s,t}^{(j)}\}_{s=1}^{t-1}\}_{t=2}^T$. Specifically, in Section \ref{subsec:inference}, we propose a BIC-based sequential procedure that automatically selects the bivariate copulas $\{\{C_{s,t}^{(j)}\}_{s=1}^{t-1}\}_{t=2}^T$ from a candidate set of parametric bivariate copulas that provide the best fit for the data.

Importantly, for the D-vine of the $j$th insurance risk, its joint distribution $f\left(y_{i1}^{(j)},\ldots,y_{iT}^{(j)}|\bm{x}_i^{(j)}\right)$ as specified in \eqref{equ:jtime} are fully characterized by the marginal count regression in \eqref{equ:mcount} and the bivariate copulas $\{\{C_{s,t}^{(j)}\}_{s=1}^{t-1}\}_{t=2}^T$ of the D-vine and can be computed efficiently via recursion. The detailed recursive evaluation procedure of the D-vine model \eqref{equ:jtime}  is summarized in Algorithm \ref{alg1} (Step I).

\textbf{The multivariate copula}: As for the conditional joint distribution of the multivariate claim counts across the $J$ perils in \eqref{equ:cjoint}, we set the $J$-variate copula $C^J$ as a Gaussian copula with an unstructured dispersion/correlation matrix. The Gaussian copula is simple to interpret, and thanks to its flexible dispersion matrix, the Gaussian copula allows different contemporaneous dependence across different pairs of insurance risks.
Based on \eqref{equ:cjoint}, simple algebra gives that, for $t=1,\cdots,T$, the joint distribution of multivariate
insurance risks for policyholder $i$ in \eqref{equ:joint} can be written as
\begin{align} \label{equ:jtyped}
	&f(\bm{y}_{it}|\bm{H}_{i,t-1})=f(y_{it}^{(1)},\cdots,y_{it}^{(J)}|H_{i,t-1}^{(1)}\cdots,H_{i,t-1}^{(J)})\nonumber\\=& \sum_{k_1=0}^{1}\cdots\sum_{k_J=0}^{1}(-1)^{k_1+\cdots+k_J}C^{J}\left(F\left(y_{it}^{(1)}-k_1|H_{i,t-1}^{(1)}\right),\ldots,F\left(y_{it}^{(J)}-k_J|H_{i,t-1}^{(J)}\right)\right),
\end{align}
with the convention $F(y|H_{i,t-1}^{(j)})=0$ for $y<0$. Note that $F(y_{it}^{(j)}-k_j|H_{i,t-1}^{(j)})$ for $k_j=0,1$ are readily available based on the evaluation of the D-vine for the $j$th insurance risk, for $j=1,\cdots,J$. This facilitates the efficient computation of the joint distribution function. The detailed procedure can be found in Algorithm \ref{alg1} (Step II).

\begin{algorithm}[]
	\begin{algorithmic}
		\State {\bf Step I}: Evaluate $f(y_{i1}^{(j)},\ldots,y_{iT}^{(j)}|\bm{x}_i^{(j)})$ by implementing steps (i)-(iii) sequentially:
		
		\hspace{-1cm}
		\begin{tabular}{cl}
		(i)&\hspace{-0.3cm}For $t=1,\ldots, T$, evaluate $F(y_{it}^{(j)}|\bm{x}_i^{(j)})$ and $f(y_{it}^{(j)}|\bm{x}_i^{(j)})$ using \eqref{equ:mcount}.\\

		(ii)&\hspace{-0.3cm}For $t=2,\ldots, T$, evaluate $f(y_{it-1}^{(j)},y_{it}^{(j)}|\bm{x}_i^{(j)})$ using (\ref{equ:paircopd}).\\
		
		(iii)&\hspace{-0.3cm}For $t=3,\ldots, T$, evaluate steps (a)-(d) for $s=t-2,t-3,\cdots,1$ recursively:\\
		\end{tabular}

~~~~(a) Calculate conditional pmf $f(y_{is}^{(j)}|y_{is+1}^{(j)},\ldots,y_{it-1}^{(j)};\bm{x}_i^{(j)})$ and\\ ~~~~~~~~~~~~~~~$f(y_{it}^{(j)}|y_{is+1}^{(j)},\ldots,y_{it-1}^{(j)};\bm{x}_i^{(j)})$ using:
\vspace{-0.2cm}
\begin{align*}
	f\left(y_{is}^{(j)}|y_{is+1}^{(j)},\ldots,y_{it-1}^{(j)};\bm{x}_i^{(j)}\right) &= \frac{f\left(y_{is}^{(j)},y_{it-1}^{(j)}|y_{is+1}^{(j)},\ldots,y_{it-2}^{(j)};\bm{x}_i^{(j)}\right)}{f\left(y_{it-1}^{(j)}|y_{is+1}^{(j)},\ldots,y_{it-2}^{(j)};\bm{x}_i^{(j)}\right)},\\
	f\left(y_{it}^{(j)}|y_{is+1}^{(j)},\ldots,y_{it-1}^{(j)};\bm{x}_i^{(j)}\right) &=
	\frac{f\left(y_{is+1}^{(j)},y_{it}^{(j)}|y_{is+2}^{(j)},\ldots,y_{it-1}^{(j)};\bm{x}_i^{(j)}\right)}{f\left(y_{is+1}^{(j)}|y_{is+2}^{(j)},\ldots,y_{it-1}^{(j)};\bm{x}_i^{(j)}\right)}.
\end{align*}
\vspace{-0.4cm}

~~~~(b) Calculate conditional cdf $F(y_{is}^{(j)}|y_{is+1}^{(j)},\ldots,y_{it-1}^{(j)};\bm{x}_i^{(j)})$ using:
\vspace{-0.2cm}
\begin{align*}
	&F\left(y_{is}^{(j)}|y_{is+1}^{(j)},\ldots,y_{it-1}^{(j)};\bm{x}_i^{(j)}\right) \\
	=& \left[C^{(j)}_{s,t-1}\left(F\left(y_{is}^{(j)}|y_{is+1}^{(j)},\ldots,y_{it-2}^{(j)};\bm{x}_i^{(j)}\right),F\left(y_{it-1}^{(j)}|y_{is+1}^{(j)},\ldots,y_{it-2}^{(j)};\bm{x}_i^{(j)}\right)\right)\right.\\
	&\left.-C^{(j)}_{s,t-1}\left(F\left(y_{is}^{(j)}|y_{is+1}^{(j)},\ldots,y_{it-2}^{(j)};\bm{x}_i^{(j)}\right),F\left(y_{it-1}^{(j)}-1|y_{is+1}^{(j)},\ldots,y_{it-2}^{(j)};\bm{x}_i^{(j)}\right)\right)\right] \Big/ \\
	&\left[F\left(y_{it-1}^{(j)}|y_{is+1}^{(j)},\ldots,y_{it-2}^{(j)};\bm{x}_i^{(j)}\right)-F\left(y_{it-1}^{(j)}-1|y_{is+1}^{(j)},\ldots,y_{it-2}^{(j)};\bm{x}_i^{(j)}\right)\right].
\end{align*}
\vspace{-0.4cm}

~~~~(c) Calculate conditional cdf $F(y_{it}^{(j)}|y_{is+1}^{(j)},\ldots,y_{it-1}^{(j)};\bm{x}_i^{(j)})$ using:
\vspace{-0.2cm}
\begin{align*}
	&F\left(y_{it}^{(j)}|y_{is+1}^{(j)},\ldots,y_{it-1}^{(j)};\bm{x}_i^{(j)}\right) \\
	=& \left[C^{(j)}_{s+1,t}\left(F\left(y_{is+1}^{(j)}|y_{is+2}^{(j)},\ldots,y_{it-1}^{(j)};\bm{x}_i^{(j)}\right),F\left(y_{it}^{(j)}|y_{is+2}^{(j)},\ldots,y_{it-1}^{(j)};\bm{x}_i^{(j)}\right)\right)\right.\\
	&\left.-C^{(j)}_{s+1,t}\left(F\left(y_{is+1}^{(j)}-1|y_{is+2}^{(j)},\ldots,y_{it-1}^{(j)};\bm{x}_i^{(j)}\right),F\left(y_{it}^{(j)}|y_{is+2}^{(j)},\ldots,y_{it-1}^{(j)};\bm{x}_i^{(j)}\right)\right)\right] \Big / \\
	&\left[F\left(y_{is+1}^{(j)}|y_{is+2}^{(j)},\ldots,y_{it-1}^{(j)};\bm{x}_i^{(j)}\right)-F\left(y_{is+1}^{(j)}-1|y_{is+2}^{(j)},\ldots,y_{it-1}^{(j)};\bm{x}_i^{(j)}\right)\right].
\end{align*}
\vspace{-0.4cm}

~~~~(d) Calculate conditional bivariate pmf $f(y_{is}^{(j)},y_{it}^{(j)}|y_{is+1}^{(j)},\ldots,y_{it-1}^{(j)};\bm{x}_i^{(j)})$ using (\ref{equ:paircopd}).\\

\State {\bf Step II}: For $t=2,\ldots, T$, evaluate $F(y_{it}^{(j)}|H^{(j)}_{i,t-1})$ using
\vspace{-0.2cm}
\begin{align*}
	F\left(y_{it}^{(j)}|H^{(j)}_{i,t-1}\right)
	=& \left[C^{(j)}_{1,t}\left(F\left(y_{i1}^{(j)}|y_{i2}^{(j)},\ldots,y_{it-1}^{(j)};\bm{x}_i^{(j)}\right),F\left(y_{it}^{(j)}|y_{i2}^{(j)},\ldots,y_{it-1}^{(j)};\bm{x}_i^{(j)}\right)\right)\right.\\
	&\left.-C^{(j)}_{1,t}\left(F\left(y_{i1}^{(j)}-1|y_{i2}^{(j)},\ldots,y_{it-1}^{(j)};\bm{x}_i^{(j)}\right),F\left(y_{it}^{(j)}|y_{i2}^{(j)},\ldots,y_{it-1}^{(j)};\bm{x}_i^{(j)}\right)\right)\right]\Big / \\
	&\left[F\left(y_{i1}^{(j)}|y_{i2}^{(j)},\ldots,y_{it-1}^{(j)};\bm{x}_i^{(j)}\right)-F\left(y_{i1}^{(j)}-1|y_{i2}^{(j)},\ldots,y_{it-1}^{(j)};\bm{x}_i^{(j)}\right)\right].
\end{align*}
\vspace{-0.4cm}


	\caption{\textcolor{black}{Evaluation of the D-vine for the $j$th insurance risk of policyholder $i$.}}
	\label{alg1}
	\end{algorithmic}
\end{algorithm}

To summarize, the D-vine based predictive model for multivariate longitudinal claim counts consists of three components. The first component is the count regression for $j=1,2,\cdots,J$, which regulates the marginal distribution of individual insurance risk given exogenous predictors. The second component is the D-vine with bivariate copulas $\{\{C^{(j)}_{s,t}\}_{s=1}^{t-1}\}_{t=2}^T$ for $j=1,2,\cdots, J$, which offers flexible modeling of temporal dependence. The last component is the $J$-variate Gaussian copula with an unstructured dispersion matrix that allows for different contemporaneous dependence among different pairs of insurance risks.

\textcolor{black}{
\textbf{Remark 2 (Extension to Cross-policyholder Dependence)}: The proposed D-vine based predictive model implicitly assumes \textit{independence} across different policyholders. This assumption may not be realistic if the insurance policies are primarily designed to cover natural catastrophes (e.g.\ extreme temperature, windstorm, hail, flood), as policyholders located in the same spatial area may be subject to the same disaster and thus exhibit contemporaneous dependence. In Section \ref{suppsec:extension} of the supplement, we provide an extension of the current model to accommodate such a scenario, where the key element is to replace the $J$-variate copula $C^J$ (that only links $J$ perils within each policyholder) with an $nJ$-variate copula $C^{*}$ that links and imposes contemporaneous dependence on $J$ perils among all $n$ policyholders. We refer to Section \ref{suppsec:extension} of the supplement for more details.
}

\subsection{Prediction of Future Risks via Stationarity}\label{subsec:prediction}
The proposed D-vine based predictive model provides a flexible framework for modeling the observed multivariate longitudinal claim counts $\{\bm{y}_{i1},\bm{y}_{i2},\cdots,\bm{y}_{iT}\}$. However, for decision making in insurance operations, the ultimate goal is prediction of future risks $\bm{y}_{i,T+1}$ given the claim history, which requires the knowledge of the conditional distribution as defined in \eqref{equ:predm}, i.e.\
\begin{align*}
	F(\bm{y}_{i,T+1}|\bm{H}_{i,T})=F\left(y^{(1)}_{i,T+1},\ldots,y^{(J)}_{i,T+1}|\bm{H}_{i,T}\right)= C^{J}\left(F\left(y_{i,T+1}^{(1)}|H_{i,T}^{(1)}\right),\ldots,F\left(y_{i,T+1}^{(J)}|H_{i,T}^{(J)}\right)\right).
\end{align*}

At first glance, it seems that $F(\bm{y}_{i,T+1}|\bm{H}_{i,T})$ may not be available as for $j=1,\dots, J$, the conditional distribution $F(y_{i,T+1}^{(j)}|H_{i,T}^{(j)})$ depends on additional bivariate copulas $\{C_{s,T+1}^{(j)}\}_{s=1}^T$, which are not specified in the D-vine for $\{y_{i1}^{(j)},\cdots,y_{iT}^{(j)}\}$. However, this issue can be naturally solved using the notion of \textit{stationarity}~\citep{Brockwell1991}, which is a fundamental assumption needed for any prediction task. In essence, stationarity of the $j$th insurance risk requires that the relationship between two observations ${y}_{is}^{(j)}$ and ${y}_{it}^{(j)}$ only depends on their distance $t-s.$ In other words, the joint distribution of $({y}_{is}^{(j)},{y}_{it}^{(j)})$ and $({y}_{is'}^{(j)},{y}_{it'}^{(j)})$ are the same if $t-s=t'-s'$.

For the $j$th D-vine to be stationary, it is easy to see that the bivariate copulas in the same tree must be the same. In other words, we require that
\begin{align}\label{equ:assum}
	C_{s,t}^{(j)}\equiv C_{s',t'}^{(j)}, \text{ for all } t-s=t'-s'.
\end{align}
Under the stationarity condition \eqref{equ:assum}, the bivariate copulas $\{C_{s,T+1}^{(j)}\}_{s=1}^T$ are readily available from the D-vine of $\{y_{i1}^{(j)},\cdots,y_{iT}^{(j)}\}$ and the predictive distribution $F(\bm{y}_{i,T+1}|\bm{H}_{i,T})$ can be efficiently computed via Algorithm \ref{alg1} by evaluating at $t=T+1.$ For the rest of the paper, we assume the stationarity condition \eqref{equ:assum} holds for the D-vine based predictive model to accommodate the prediction-oriented nature of our study for bundled insurance risks. 

\section{Statistical Estimation and Inference} \label{sec:inference}
The D-vine based predictive model proposed in Section \ref{sec:method} is of parametric nature and we thus design likelihood-based methods for its estimation and inference. For simplicity, with a slight abuse of notation, for risk $j=1,2,\cdots,J,$ we use $\bfbeta_j$ to denote all parameters involved in the marginal count regression, use $\bfzeta_j$ to denote all parameters involved in the bivariate copulas $\{\{C_{s,t}^{(j)}\}_{s=1}^{t-1}\}_{t=2}^T$ of the D-vine, and we use $\bfrho$ to denote parameters in the Gaussian copula $C^J.$ Furthermore, denote $\bfbeta=(\bfbeta_1,\cdots,\bfbeta_J)$, denote $\bfzeta=(\bfzeta_1,\cdots,\bfzeta_J)$, and collect all model parameters as $\bftheta=(\bfbeta,\bfzeta,\bfrho)$.

Section \ref{subsec:est} discusses a three-stage sequential maximum likelihood estimator~(MLE) for the estimation of $\bftheta$ and establishes its theoretical guarantees. Section \ref{subsec:inference} further proposes a sequential model selection procedure that automatically selects the bivariate copulas $\{\{C_{s,t}^{(j)}\}_{s=1}^{t-1}\}_{t=2}^T$ from a candidate set of bivariate parametric copulas in a fully data-driven fashion.

\subsection{Parameter Estimation}\label{subsec:est}
Based on joint distribution of multivariate insurance risks in \eqref{equ:joint}, given a portfolio of $n$ policyholders observed for $T$ periods $\{(\bm{y}_{i1},\ldots,\bm{y}_{iT})\}_{i=1}^n$, the full log-likelihood function can be written as
\begin{align} \label{equ:loglik}
L(\bm{\theta}) = L(\bfbeta,\bfzeta,\bfrho) = \sum_{i=1}^{n}\sum_{t=1}^{T}\log f(\bm{y}_{it}|\bm{H}_{i,t-1})=\sum_{i=1}^{n} l_i(\bftheta),
\end{align}
where we denote $l_i(\bftheta)=\sum_{t=1}^{T}\log f(\bm{y}_{it}|\bm{H}_{i,t-1}).$

In principle, we can estimate $\bftheta$ by directly maximizing $L(\bftheta)$. However, due to the discreteness of the insurance claim counts $\{(\bm{y}_{i1},\ldots,\bm{y}_{iT})\}_{i=1}^n$, the evaluation of $L(\bftheta)$ can be computationally expensive, making the joint estimation of $(\bfbeta,\bfzeta,\bfrho)$ infeasible when the portfolio of insurance contracts is of large size $n$, a commonly encountered situation for insurers in the ear of big data. Thus, for computational efficiency, we instead propose a three-stage sequential estimation procedure in the same spirit of inference function for margins (IFM)~\citep{Joe2005}. In particular, we employ a divide-and-conquer strategy and estimate $\bfbeta$, $\bfzeta$, and $\bfrho$ sequentially one by one.

\textbf{Three-stage MLE}: In the first stage, we focus on the estimation of the marginal count regression parameter $\bfbeta$. For computational efficiency, we (purposely) assume temporal and contemporaneous independence among the multivariate insurance risks. Under the imposed working independence assumption, the full log-likelihood function in \eqref{equ:loglik} simplifies to
\begin{align*}
	L_1(\bfbeta)=L_1(\bfbeta_1,\cdots,\bfbeta_J)=\sum_{j=1}^J\left(\sum_{i=1}^{n}\sum_{t=1}^{T}\log f({y}^{(j)}_{it}|\bm{x}_{i}^{(j)})\right)=\sum_{j=1}^{J} \left(\sum_{i=1}^{n} l_{1i}^{(j)}(\bfbeta_j)\right),
\end{align*}
where we denote $l_{1i}^{(j)}(\bfbeta_j)=\sum_{t=1}^{T}\log f({y}^{(j)}_{it}|\bm{x}_{i}^{(j)})$ and $f({y}^{(j)}_{it}|\bm{x}_{i}^{(j)})$ is the marginal likelihood specified in \eqref{equ:mcount}. Define $L_{1}^{(j)}(\bfbeta_j)=\sum_{i=1}^{n}l_{1i}^{(j)}(\bfbeta_j)$, the estimator $\widehat{\bfbeta}=(\widehat{\bfbeta}_1,\cdots,\widehat{\bfbeta}_J)$ can be efficiently obtained via $J$ separate optimization
\begin{align*}
	\widehat{\bfbeta}_j=\argmax L_{1}^{(j)}(\bfbeta_j), \text{ for } j=1,\cdots, J.
\end{align*}

In the second stage, we focus on the estimation of the D-vine bivariate copula parameter $\bfzeta.$ To reduce computational complexity, we (purposely) fix $\bfbeta=\widehat{\bfbeta}$ (as estimated in the first stage) and (purposely) assume contemporaneous independence across the $J$ insurance risks. In other words, we set the multivariate copula $C_J$ as the independence copula, which indicates that $f(\bm{y}_{it}|\bm{H}_{i,t-1})$ in \eqref{equ:jtyped} simplifies to
$f(\bm{y}_{it}|\bm{H}_{i,t-1})= \prod_{j=1}^J(F(y_{it}^{(j)}|H_{i,t-1}^{(j)})-F(y_{it}^{(j)}-1|H_{i,t-1}^{(j)}))$. The full log-likelihood function in \eqref{equ:loglik} is thus simplified to
\begin{align*}
	&L_2(\bfzeta)=L_2(\bfzeta_1,\cdots,\bfzeta_J)\\
	=&\sum_{j=1}^J\left(\sum_{i=1}^{n}\sum_{t=1}^{T}\log \left(F(y_{it}^{(j)}|H_{i,t-1}^{(j)})-F(y_{it}^{(j)}-1|H_{i,t-1}^{(j)})\right)\right)=\sum_{j=1}^{J} \left(\sum_{i=1}^{n} l_{2i}^{(j)}(\bfzeta_j;\widehat{\bfbeta}_j)\right),
\end{align*}
where we denote $l_{2i}^{(j)}(\bfzeta_j; \widehat{\bfbeta}_j)=\sum_{t=1}^{T}\log (F(y_{it}^{(j)}|H_{i,t-1}^{(j)})-F(y_{it}^{(j)}-1|H_{i,t-1}^{(j)}))$. Define $L_{2}^{(j)}(\bfzeta_j)=\sum_{i=1}^{n}l_{2i}^{(j)}(\bfzeta_j;\widehat{\bfbeta}_j)$, the estimator $\widehat{\bfzeta}$ can again be efficiently obtained via $J$ separate optimization
\begin{align*}
	\widehat{\bfzeta}_j=\argmax L_{2}^{(j)}(\bfzeta_j), \text{ for } j=1,\cdots, J.
\end{align*}

In the last stage, we estimate the contemporaneous dependence parameter $\bfrho$ of copula $C_J$ while fixing $\bfbeta=\widehat{\bfbeta}$ and $\bfzeta=\widehat{\bfzeta}$ in the full log-likelihood function \eqref{equ:loglik}. In other words, we obtain $\widehat{\bfrho}$ via
\begin{align*}
	\widehat{\bfrho}=\argmax L(\widehat{\bfbeta},\widehat{\bfzeta},\bfrho).
\end{align*}

Compared to the classical MLE, the proposed three-stage MLE $\widehat{\bftheta}=(\widehat{\bfbeta},\widehat{\bfzeta},\widehat{\bfrho})$ decomposes the joint estimation of a potentially large parameter vector $\bftheta$ into three separate estimation problems and thus achieves substantial computational efficiency, enabling fast implementation of the sophisticated D-vine based predictive model for large-scale insurance data.

\textbf{Theoretical guarantees}: To conserve space, we present the detailed theoretical guarantees for $\widehat{\bftheta}$ in Section \ref{sec:thm}~(Theorem \ref{thm:mle}) of the supplement. In short, same as the classical MLE, the three-stage MLE $\widehat{\bftheta}$ is consistent, asymptotically normal, and admits an asymptotic error of order $O_p(1/\sqrt{n})$. On the other hand, due to the (incorrect) working independence assumption and the sequential nature of the three-stage estimation procedure, which are purposely employed for computational efficiency, the asymptotic covariance of $\widehat{\bftheta}$ admits a more complicated Godambe form~\citep{Godambe1960} and is statistically less efficient than the classical MLE. Intuitively, the working independence assumption incurs inefficiency due to model mis-specification and the sequential estimation incurs inefficiency as the estimation error in previous stages affects the estimation in the subsequent stages. We refer to \cite{Newey1994} for more discussion of such phenomenon.

In other words, there is a trade-off between computational and statistical efficiency. However, the concern of statistical inefficiency can be well addressed when the insurance portfolio size $n$ is large, which is common for modern large-scale insurance data. In contrast, for large-scale data, the computational inefficiency is the more prominent concern.

Though the standard plug-in estimator can be constructed for the asymptotic covariance of $\widehat{\bftheta}$, it can be quite cumbersome (both analytically and numerically) to implement as the asymptotic covariance~(see Theorem \ref{thm:mle} of the supplementary material) does not admit a closed form due to its complexity as a result of the three-stage estimation. A practical solution to the estimation of the asymptotic covariance is parametric bootstrap, see for example \cite{ZhaoZhang2017}.


\textbf{Numerical results}: We conduct extensive simulation experiments to examine finite-sample performance of the proposed three-stage MLE $\widehat{\bftheta}$ under simulation settings that resemble real data used in our empirical analysis. To conserve space, we present the detailed results in Section \ref{subsec:simu_paraest} of the supplementary material. In summary, the three-stage MLE $\widehat{\bftheta}$ achieves satisfactory estimation accuracy and the confidence interval of $\widehat{\bftheta}$ constructed based on the parametric bootstrap provides adequate coverage rates under portfolio size as small as $n=500$.

\subsection{Data-driven Bivariate Copula Selection in D-vine}\label{subsec:inference}
As discussed in Section \ref{subsec:vine_framework}, an attractive feature of the D-vine is its flexibility, as different combinations of bivariate copulas $\{\{C_{s,t}^{(j)}\}_{s=1}^{t-1}\}_{t=2}^T$ offer substantial potential to accommodate a wide range of temporal dependence for the $j$th insurance risk with $j=1,2,\cdots, J.$ However, in practice, the optimal specification of $\{\{C_{s,t}^{(j)}\}_{s=1}^{t-1}\}_{t=2}^T$ is unknown and we thus propose a data-driven procedure that automatically selects $\{\{C_{s,t}^{(j)}\}_{s=1}^{t-1}\}_{t=2}^T$ that provide the best fit for the given data.

Given a set of candidate bivariate copulas~(say $m$ different copulas), under the stationarity condition \eqref{equ:assum}, the number of possible combinations of $\{\{C_{s,t}^{(j)}\}_{s=1}^{t-1}\}_{t=2}^T$ is $m^{T-1}$, as there are $T-1$ trees of the D-vine and all bivariate copulas in the same tree are the same. Note that $m^{T-1}$ can be quite large even for moderate $m$ and $T$. Thus, for computational feasibility, we propose a BIC-based tree-by-tree sequential selection procedure. As is evident from Algorithm \ref{alg1}, the evaluation of higher level trees of a D-vine relies on the lower level trees, thus, our selection procedure sequentially traverses from tree 1 to tree $T-1.$

The basic procedure is as follows. We start with the bivariate copula in tree 1, selecting the suitable copula from a given set of candidates and estimating its model parameter. Fixing the selected copula and estimated parameter in the first tree, we then select the optimal copula and estimate its parameter for the second tree. We continue this process for the next tree of a higher level while holding the selected copulas and the corresponding estimated parameters fixed in all previous trees. If the bivariate copula is selected as the independence copula for a certain tree, we then stop the selection procedure and truncate the D-vine, i.e. select independence copulas for all higher level trees (see, for example, \cite{BrechmannCzadoAas2012}).

This truncation mechanism has an important implication for the predictive modeling application. A key component of accurate prediction for future risks is to determine the optimal credibility weight given to historical measurements of the insurance risks, which is similar to order selection of an auto-regressive model in time series analysis~\citep{Brockwell1991}. This issue is nicely addressed by the truncation mechanism in a data-driven fashion, since if the D-vine is truncated at tree $k$, the prediction of future risk at time $T+1$ will only depend on the claim history from time $T-k+1$ to time $T$. In other words, the selected D-vine deems the historical risk measurements observed prior to time $T-k+1$ irrelevant for prediction purposes.

\begin{figure}[htp]
  \begin{center}
   \includegraphics[width=0.97\textwidth,angle=0]{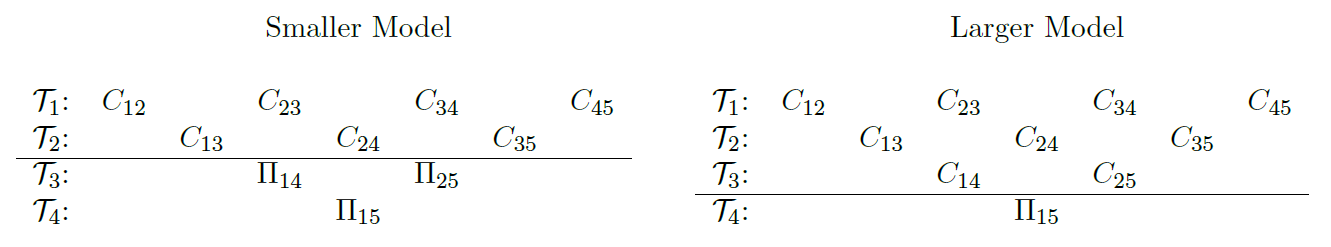}
  \end{center}
  \vspace{-0.8cm}
   \caption{Graphical representation of truncated 5-dimensional D-vines.}
   \label{fig:tdvine}
\end{figure}

Figure \ref{fig:tdvine} provides two examples of truncated 5-dimensional D-vines, where $C$ denotes selected non-independence bivariate copulas and $\Pi$ denotes the independence copula. The left panel corresponds to a smaller model where the D-vine is truncated at the second tree, and the right panel shows a larger model where truncation occurs at the third tree.

The commonly used BIC is employed for bivariate copula selection in each tree. Specifically, given the $j$th insurance risk, for tree $k$ of the D-vine, we define the BIC-based model selection criterion for a candidate bivariate copula $C$ as
\begin{align*} 
	\text{BIC}_{k}^{(j)} = -2\sum_{i=1}^n \sum_{ \substack{1\leq s\leq T-k\\ t=s+k}} \log f\left(y^{(j)}_{is},y^{(j)}_{it}|y^{(j)}_{is+1},\ldots,y^{(j)}_{it-1};\bm{x}_i^{(j)}\right) + \log n\cdot(\text{number of parameters in } C),
\end{align*}
where the likelihood function $f(y^{(j)}_{is},y^{(j)}_{it}|y^{(j)}_{is+1},\ldots,y^{(j)}_{it-1};\bm{x}_i^{(j)})$ can be readily evaluated based on the selected copulas $\{C_{s,t}^{(j)}\}_{t-s<k}$ of previous trees~(i.e.\ tree 1 to tree $k-1$) and the current candidate copula $C$ for tree $k$~(see Algorithm \ref{alg1} for more details). We select the bivariate copula in the candidate set that minimizes $\text{BIC}_{k}^{(j)}$.

We examine the performance of the sequential tree-by-tree selection procedure via extensive numerical experiments in Section \ref{subsec:simu_copselection} of the supplement, where it is seen to be computationally efficient and can accurately identify the true truncation order of the D-vine~(i.e.\ the optimal weight given to historical observations for prediction) and the true bivariate copula for each tree.

\section{Empirical Analysis}\label{sec:empirical}
\subsection{Data Description}
Our dataset features bundled insurance risks from a multi-peril property insurance fund in the state of Wisconsin. The fund is established to provide property insurance for local government entities including counties, cities, towns, villages, school districts, fire departments, and other miscellaneous entities, and is administered by the Wisconsin Office of the Insurance Commissioner. 
The fund operates and functions as a stand-alone commercial property insurer in that it charges premiums and pays claims to its policyholders, i.e. local government units. On average, the fund writes approximately \$25 million in premiums and \$75 billion in coverage each year.

One major coverage that the fund provides to local government entities is building and contents, where the building element covers the physical structure of a property including its permanent fixtures and fittings, and the contents element covers possessions and valuables within the property that are detached and removable. It is an open-peril policy such that the policy insures against loss to covered properties from all causes with certain exclusions. Such exclusions include those resulting from flood, earthquake, wear and tear, extremes in temperature, mold, war, nuclear reactions, and embezzlement or theft by an employee. The fund groups all perils into three categories: water, fire, and others. Each category represents an insurance risk and there are $J=3$ bundled risks associated with a single insurance policy. The measurement of insurance risks is the number of claims per year. The dataset contains detailed policyholder-level information, including exogenous predictors and claim history, for 1019 local government entities over a six-year period from 2006-2011. 

To conserve space, we provide detailed description of the dataset, including descriptive statistics of the exogenous predictors and the multivariate longitudinal claim counts, and sample dependence measures, in Section \ref{sec:moreresults} of the supplementary material. In the empirical analysis conducted in Sections \ref{sec:empirical}-\ref{sec:manage}, we use the observations in years 2006-2010 as the training data to develop the D-vine based predictive model~(i.e.\ we have $n=1019$ and $T=5$ for model estimation) and use the data in year 2011 as the hold-out sample for model validation and comparison.



\subsection{Model Specification and Estimation}\label{subsec:model_est}
To analyze the claim count of each peril (water, fire, and other), we consider the zero-one inflated negative binomial regression \eqref{equ:mcount} proposed in Section \ref{subsec:riskmodel}. In addition, we examine its nested and limiting cases, including standard Poisson and negative binomial regression, and the count regression models with inflation only in zero and one respectively.

To assess the goodness-of-fit, we compare the empirical claim frequency with the frequency implied from the fitted regression model, which is calculated as the sum of the estimated probabilities of claim count over all policyholders at each observed value. The count regression model is then selected for each peril based on the chi-squared statistics. We refer to Table \ref{tab:count} in Section \ref{sec:moreresults} of the supplementary material for more details on the empirical claim frequency of each peril. Table \ref{tab:countfit} reports the goodness-of-fit statistics for the selected model as well as the alternative candidates. For the peril of water, fire, and others, we select the zero-one inflated negative binomial model (ZOINB), standard negative binomial model (NB), and one inflated negative binomial model (OINB), respectively.

The claim counts from all perils exhibit strong evidence of overdispersion that cannot be well-captured by Poisson-based models, a feature commonly observed in insurance claim data. Since insurance operation is based on risk pooling, the overdispersion in claim counts is usually attributed to the excess of zeros corresponding to the large number of policyholders without any claims. 
An interesting finding for the peril-wise claim count data in our study is, in addition to the zero inflation, there is a significant portion of ones as evidenced by the selected count regression models for the water and other perils.

\begin{table}[htbp]
  \centering
  \caption{Goodness-of-fit statistics of different count regression models for claim count by peril. Z,O,I,P,NB stands for zero, one, inflated, Poisson and negative binomial, respectively.}
    \begin{tabular}{llr|rrr}
    \hline\hline
              & \multicolumn{2}{c|} {}   & \multicolumn{3}{c} {Alternative Models} \\
     Peril      & \multicolumn{2}{c|} {Selected Model}         &      Poisson & ZIP & ZOIP  \\
    \hline
    Water &  ZOINB   & 24.604  & 151.139  & 61.891 & 39.489 \\
    Fire &   NB    & 8.178   & 63.953 &  8.945  & 8.945\\
    Other &  OINB     & 26.734 & 100.393 & 81.211 & 48.520   \\
  \hline\hline
    \end{tabular}%
  \label{tab:countfit}%
\end{table}%

The estimation results for the selected count regression models are given in Table \ref{tab:countest}. We refer to Table \ref{tab:covariates} in Section \ref{sec:moreresults} of the supplementary material for more details on the exogenous predictors used in the regression. Some observations are as follows. There exists significant difference in claim frequency across different entity types~(City, County, School, Town, Village), though the effects are heterogeneous across the three perils. The alarm credit~(AC05, AC10, AC15) is not predictive regardless of the peril type, which is intuitively understandable as the alarm system is a more effective tool for loss control rather than loss prevention. The amount of coverage of the insurance policy shows substantial predictive power for all perils, indicating it is a sensible measure of risk exposure for the policyholder.

\begin{table}[htbp]
  \centering
  \caption{Estimation results of count regression models for claim count by peril. For exogenous predictors: (a) City, County, School, Town, Village are entity type indicators; (b) AC05, AC10, AC15 are alarm credit indicators; (c) Coverage is the amount of coverage of the insurance policy.}
    \begin{tabular}{lrrrrrrrr}
    \hline\hline
          & \multicolumn{2}{c}{Water}  &       & \multicolumn{2}{c}{Fire}     &       & \multicolumn{2}{c}{Other}  \\
    \cline{2-9}
          & EST. & S.E. &       & EST. & S.E &       & EST. & S.E \\
    \hline
    Intercept & -5.985 & 0.365 &       & -3.987 & 0.229 &       & -5.276 & 0.401 \\
    City & 1.270 & 0.318 &       & 1.072 & 0.216 &       & 0.739 & 0.340 \\
    County & 0.570 & 0.341 &       & 1.622 & 0.223 &       & 0.974 & 0.350 \\
    School & -0.273 & 0.321 &       & 0.170 & 0.217 &       & 0.299 & 0.338 \\
    Town & 1.767 & 0.450 &       & 0.093 & 0.326 &       & 0.325 & 0.593 \\
    Village & 1.381 & 0.332 &       & 1.047 & 0.221 &       & 0.690 & 0.364 \\
    AC05  & -0.166 & 0.338 &       & 0.098 & 0.228 &       & 0.324 & 0.312 \\
    AC10  & -0.099 & 0.273 &       & 0.276 & 0.179 &       & 0.107 & 0.285 \\
    AC15  & 0.065 & 0.144 &       & 0.141 & 0.102 &       & 0.086 & 0.158 \\
    Coverage & 1.225 & 0.061 &       & 0.477 & 0.036 &       & 0.808 & 0.057 \\
    $\phi$ & 0.279 & 0.044 &       & 1.142 & 0.177 &       & 0.370 & 0.062 \\
    \hline
    Zero Model &       &       &       &       &       &       &       &  \\
    Intercept & -4.175 & 1.452 &       &       &       &       &       &  \\
    Coverage & 0.495 & 0.204 &       &       &       &       &       &  \\
    \hline
    One Model  &       &       &       &       &       &       &       &  \\
    Intercept & -3.356 & 0.152 &       &       &       &       & -4.254 & 0.268 \\
    Coverage & 0.159 & 0.055 &       &       &       &       & 0.316 & 0.072 \\
  \hline\hline
    \end{tabular}%
  \label{tab:countest}%
\end{table}%

For the dependence model, we employ the pair copula construction with D-vine to accommodate the temporal dependence for longitudinal measurements of each peril. The tree-by-tree sequential selection procedure proposed in Section \ref{subsec:inference} is used to automatically select the bivariate copulas for each D-vine, and we consider a candidate set of bivariate copulas that contains the most widely used copulas in practice, including the Independence, Gaussian, Frank, (rotated) Clayton, (rotated) Gumbel, and (rotated) Joe copulas. Given the selected bivariate copulas for each D-vine, we estimate the model parameters using the three-stage MLE described in Section \ref{subsec:est}.


The selected bivariate copulas and the estimated model parameters for the three D-vines are reported in Table \ref{tab:countdvine}. To better gauge the magnitude of the dependence, we further present the Kendall's $\tau$ implied by the bivariate copulas. Some observations are as follows. First, a wide range of bivariate copulas are selected for each D-vine, which confirms that D-vine can accommodate flexible temporal dependence. Second, the estimated temporal dependence is positive, implying the imperfection of the insurer's risk classification system. This further foreshadows the important role of the predictive model in improving the insurance operations~(Section \ref{sec:manage}). Third, within each D-vine, the estimated dependence of bivariate copulas decreases from the top to the bottom trees. The diminishing dependence pattern is consistent with the fundamental idea of the graphical model in that pairs in higher level trees are conditioned on a larger set of correlated variables, and thus are expected to be less dependent. In particular, note that the D-vines for both water and other perils are truncated~(i.e.\ conditional independence beyond a certain tree level), with the former at the third tree and the latter at the second tree, indicating that the optimal credibility weights given to historical claim counts data are different across the three perils.


\begin{table}[htbp]
  \centering
  \caption{Selected bivariate copulas and its estimated model parameters (Est.) and Kendall's $\tau$ in each D-vine by peril. Standard errors are presented in parenthesis. R.\ stands for Rotated. $T_k$ denotes the $k$th tree for $k=1,\cdots,4$.}
  \begin{tabular}{llccclccclcc}
    \hline\hline
          & \multicolumn{3}{c}{Water} & &    \multicolumn{3}{c}{Fire} &  & \multicolumn{3}{c}{Other}  \\
          \cline{2-4}\cline{6-8}\cline{10-12}
          & Copula & Est. & $\tau$ && Copula & Est. & $\tau$ && Copula & Est. & $\tau$ \\
    \hline
    $T_1$ & {R.Gumbel } & 1.547 & 0.359 & & {R.Gumbel} & 1.299 & 0.222 & & {Clayton} & 0.920 & 0.315 \\
          &       & (0.054) &      & &  &(0.045) & &   &   & (0.170) & \\
    $T_2$ & {R.Gumbel} & 1.321 & 0.246 & & {R.Gumbel } & 1.312 & 0.231 && {R.Gumbel } & 1.252 & 0.201 \\
          &       & (0.053) &       &&   & (0.052) &  & &     & (0.057) & \\
    $T_3$ & {Frank} & 1.529 & 0.166 && {R.Gumbel } &  1.216 & 0.179 &      &  &&\\
          &       & (0.342) & &&       & (0.056) & &       & && \\
    $T_4$ &       &        && & {R.Clayton }  & 0.172 & 0.080 &       &  &&\\
          &       &        &&  &     & (0.052) & &       &  &&\\
    \hline\hline
  \end{tabular}%
  \label{tab:countdvine}%
\end{table}%

The contemporaneous dependence among the three perils are captured using a Gaussian copula with an unstructured dispersion matrix. The estimated pairwise association parameters are reported in Table \ref{tab:paircorr}, where significant dependence is observed among the three types of risks. The results support the unstructured association across claim counts from different perils and are consistent with the observations in Figure \ref{fig:corcount} of the supplementary material. We remark that the proposed framework  allows alternative copula specifications of $C_J$ to capture the contemporaneous dependence among the multivariate insurance risks, such as factor copulas and hierarchical Archimedean copulas.
In our current study, the Gaussian copula is a favorable choice due to its balance between interpretability and computational difficulty. Moreover, our analysis shows that the Gaussian copula sufficiently captures the association among insurance claim counts of different perils.

\begin{table}[htbp]
  \centering
  \caption{Estimates of pairwise contemporaneous dependence among perils}
    \begin{tabular}{lccc}
    \hline\hline
    & Water-Fire       & Water-Other        & Fire-Other\\
    \hline
Estimate   & 0.126 & 0.199 & 0.070 \\
$t$-stat & 3.558 & 5.083 & 1.887 \\
    \hline\hline
    \end{tabular}%
  \label{tab:paircorr}%
\end{table}%

\subsection{Model Validation}
Business decisions in insurance operations are often based on the forecast of risk outcomes in the future period. To reflect the uncertainty of the future, any prediction should be treated as probabilistic, i.e.\ they should take form of probability distributions over future quantities. Hence, we perform model validation for the predictive distribution (\ref{equ:predm}) instead of point prediction, and we examine the predictive performance of the proposed model using the proper scoring rules for probabilistic forecasts~\citep{Gneiting2007,CzadoGneitingHeld2009}.

Our validation procedure is based on the aggregate risk $S_{i,T+1}$ defined in \eqref{equ:indagg}. Without loss of generality, we set $\alpha_{i,T+1}^{(j)}=1$ for $j\in\{1,\ldots,J\}$ and $i \in \{1,\ldots,n\}$ for simplicity. Thus $S_{i,T+1}$ is the total number of claims from all perils for the $i$th policyholder in the future period $T+1$~(i.e.\ year 2011). Let $F_S(\cdot|\bm{H}_{iT})$ denote the predictive distribution of $S_{i,T+1}$ conditional on history $\bm{H}_{iT}$, i.e.,
\begin{align*}
F_S(s|\bm{H}_{iT}) & = {\rm Pr}(S_{i,T+1}\le s|\bm{H}_{iT}) =\sum_{\left\{\left(y_{i,T+1}^{(1)},\cdots,y_{i,T+1}^{(J)}\right)\middle| S_{i,T+1}\le s\right\}} f\left(y^{(1)}_{i,T+1},\ldots,y^{(J)}_{i,T+1}|H_{iT}^{(1)},\ldots,H_{iT}^{(J)}\right).
\end{align*}
Note that $f(y^{(1)}_{i,T+1},\ldots,y^{(J)}_{i,T+1}|H_{iT}^{(1)},\ldots,H_{iT}^{(J)})$ can be obtained from the predictive model \eqref{equ:predm} as
\begin{align*}
	& f\left(y^{(1)}_{i,T+1},\ldots,y^{(J)}_{i,T+1}|H_{iT}^{(1)},\ldots,H_{iT}^{(J)}\right)\nonumber \\
	= &\sum_{k_1=0}^{1}\cdots\sum_{k_J=0}^{1}(-1)^{k_1+\cdots+k_J} F\left(y^{(1)}_{i,T+1}-k_1,\ldots,y^{(J)}_{i,T+1}-k_J|H_{iT}^{(1)},\ldots,H_{iT}^{(J)}\right),
\end{align*}
and as discussed in Section \ref{subsec:prediction}, $F(y^{(1)}_{i,T+1},\ldots,y^{(J)}_{i,T+1}|H_{iT}^{(1)},\ldots,H_{iT}^{(J)})=F(\bm{y}_{i,T+1}|\bm{H}_{i,T})$ and can be evaluated efficiently via Algorithm \ref{alg1} by setting $t=T+1$. 

Given the predictive distribution $\{F_S(\cdot|\bm{H}_{iT})\}_{i=1}^n$ given by the model, the validation is based on the probability integral transformation (PIT) of $\{S_{i,T+1}\}_{i=1}^n$. It is well-known that for a continuous random variable $Z$ with cdf $F_Z(\cdot)$, the PIT $F_Z(Z)$ is a uniform random variable. For the portfolio of policyholders, denote $s_{i,T+1}$ as the realized value of $S_{i,T+1}$ and define $u_{i,T+1}=F_S(s_{i,T+1}|\bm{H}_{iT})$. Therefore, the validity of the predictive model can be empirically tested by examining whether $\{u_{i,T+1};i=1,\cdots,n\}$ is a random sample from the uniform distribution.

\begin{figure}[htp]
  \begin{center}
  \includegraphics[width=0.38\textwidth,angle=0]{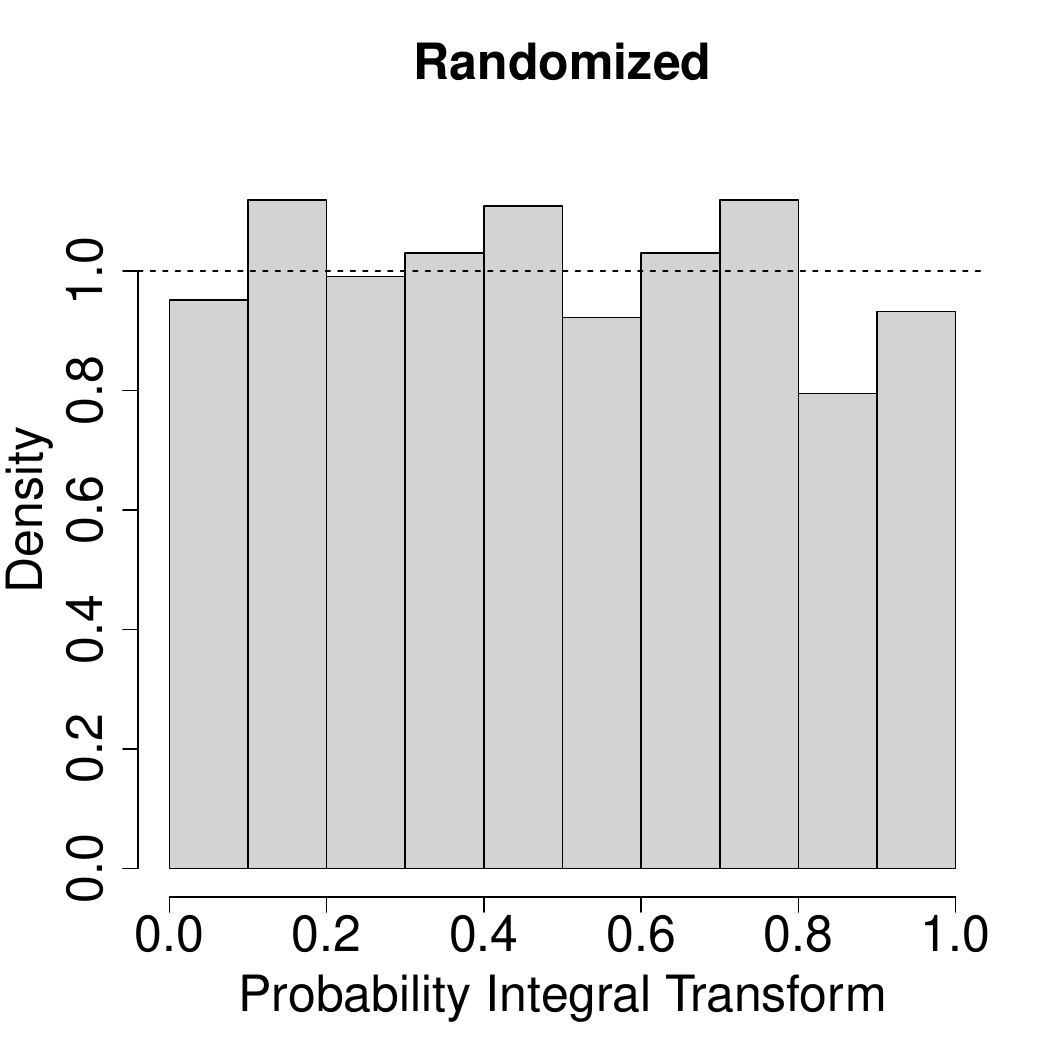}
  \includegraphics[width=0.38\textwidth,angle=0]{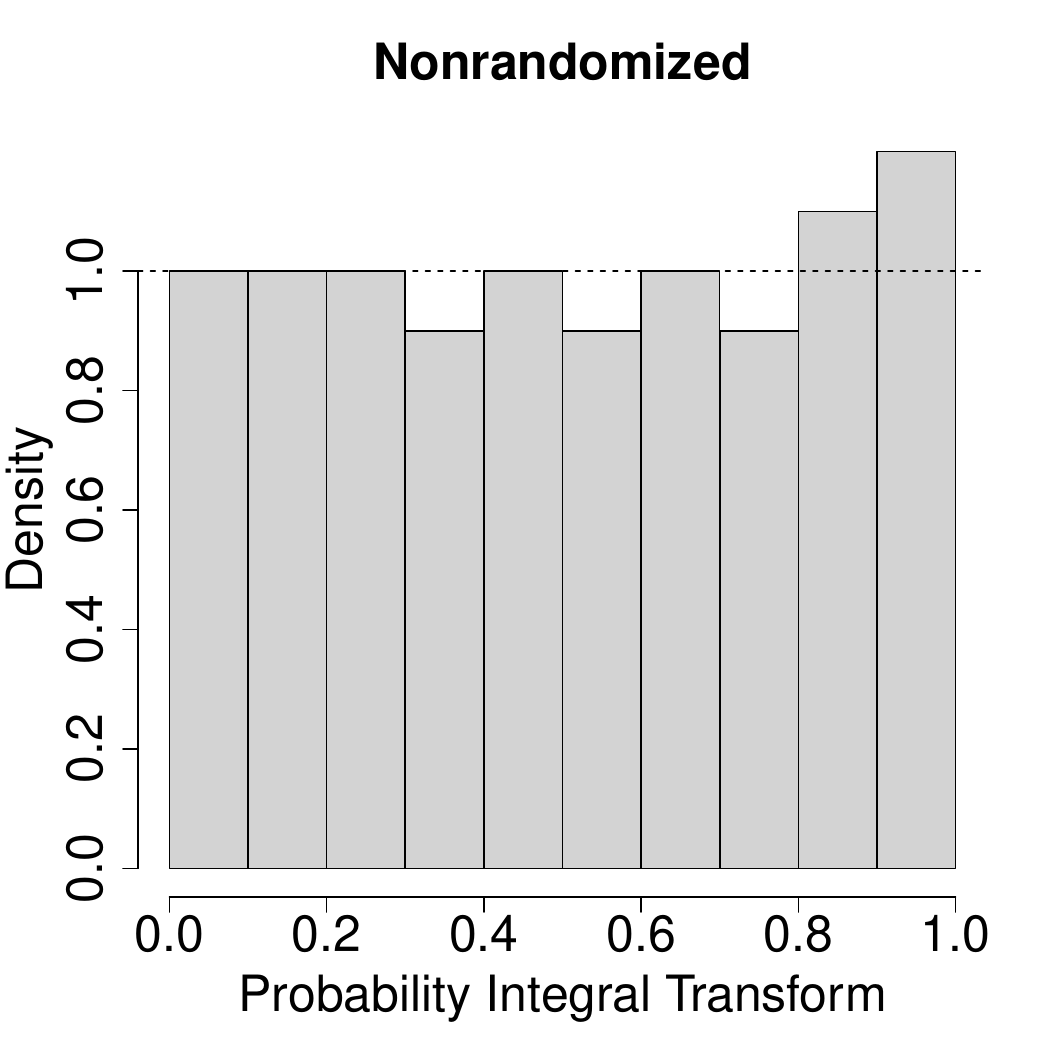}
  \end{center}
  \vspace{-0.8cm}
  \caption{PIT histograms based on the predictive distribution.}
  \label{fig:pit}
\end{figure}

However, in our context, $S_{i,T+1}$ represents the number of claims at the policy level. Due to the discrete nature of $S_{i,T+1}$, the standard PIT is not applicable. To address this issue, we instead consider a randomized PIT defined by (see \cite{Ruschendorf2009}):
\begin{align*}
U_{i} =  F_S(S_{i,T+1}-1|\bm{H}_{iT}) + V_i\cdot (F_S(S_{i,T+1}|\bm{H}_{iT}) - F_S(S_{i,T+1}-1|\bm{H}_{iT})),
\end{align*}
where $V_i$ is \textit{i.i.d.}\ uniform random variable and we define $F_S(-1|\bm{H}_{iT})=0$ for $i=1,\cdots,n$. In addition, we consider a nonrandomized PIT which is defined based on the conditional CDF of $U_i$ given the observed claim count $S_{i,T+1}=s_{i,T+1}$ (see Section 2.1 in \cite{CzadoGneitingHeld2009} for more details). The randomized and nonrandomized PIT histograms are exhibited in Figure \ref{fig:pit}. Both transformations are uniformly distributed, suggesting the predictive distribution based on the proposed D-vine model is well calibrated. The uniformity is formally tested using the Kolmogorov-Smirnov~(KS) statistics for both randomized and nonrandomized PITs. For the randomized PIT, the KS statistic is 0.0307, which gives a $p$-value of 0.294. For the nonrandomized PIT, the KS statistic is 0.0326, which gives a $p$-value of 0.199. The result provides further support that the predictive distribution based on the D-vine model is sufficiently flexible for insurance applications. 

To further illustrate the importance of dependence modeling for prediction, we further compare the predictive accuracy between the proposed D-vine model and the independence model, which ignores the temporal and contemporaneous dependence among multivariate bundled insurance risks.

The assessment is again based on the predictive distribution of $S_{i,T+1}$. We employ consistent scoring rules to evaluate the sharpness/accuracy of the predictive distribution estimated by the two models. Specifically, we consider three scoring rules introduced in \cite{CzadoGneitingHeld2009}, the ranked probability score (RPS), the quadratic score (QS), and the spherical score (SPHS), which essentially quantify the closeness between the realized value $s_{i,T+1}$ in the hold-out sample and the predictive distribution $F_S(\cdot|\bm{H}_{iT})$ estimated based on training data. All scores are negatively oriented, i.e., the lower the score, the more accurate the prediction. The three scores are calculated for each of the 1019 policyholders for both the independence model and the D-vine model. Table \ref{tab:predscore} reports the empirical probability that the D-vine model outperforms the independence model. The dependence-aware prediction based on the D-vine model is superior to the independence model for about 70\% of policyholders in the hold-out sample across all three scoring rules. The large $Z$-scores for the one-sided binomial tests further confirm the statistical significance of the result.

\begin{table}[htbp]
  \centering
  \caption{Empirical probability of superior prediction by the D-vine model}
    \begin{tabular}{lrrr}
\hline\hline
          & RPS &   QS    & SPHS \\
\hline
  Estimate        & 69.09\% &   68.40\%    & 68.99\% \\
 $Z$-score        & 12.19 &   11.74    & 12.12 \\
\hline\hline
    \end{tabular}%
  \label{tab:predscore}%
\end{table}%

\section{Managerial Implications on Key Insurance Operations}\label{sec:manage}
This section illustrates in detail the prominent managerial implications of the proposed D-vine predictive model, in particular, the dependence among multivariate longitudinal insurance risks, on several key insurance operations outlined in Section \ref{sec:generalsetting}.

We emphasize that current modeling practice in the insurance industry mostly generates prediction for bundled insurance risks without accounting for any dependence or only accounts for temporal dependence. In contrast, the D-vine predictive model enables simultaneous analysis of both temporal and contemporaneous dependence, and thus generates fully dependence-aware prediction. We show below that the dependence-aware prediction significantly improves the insurer's profitability in risk pricing and provides more accurate risk assessment for reinsurance.

We consider the standard scenario where business decisions are made based on the prediction of future aggregate loss of individual policyholders. Specifically, given historical information on an insurance portfolio over the past $T$ years, the insurer is to make operational decisions based on the prediction for year $T+1$. A policyholder's aggregate loss in year $T+1$ is defined in \eqref{equ:indagg} and we derive the average cost per claim via a generalized linear model (GLM). Specifically, we fit a Gamma GLM using data in the first $T$ years, where the response variable is the average cost per claim and the predictors are summarized in Table \ref{tab:covariates} of Section \ref{sec:moreresults} of the supplementary material. See \cite{DeJongHeller2008} for implementation of GLMs for insurance data. A GLM is fitted separately for each peril, and the predicted outcome for year $T+1$ is used as the average cost per claim in the following analysis.



\subsection{Risk Segmentation and Pricing} \label{subsec:ratemaking}
We first examine the insurer's underwriting and ratemaking practice. To identify profitable business, it is essential for the insurer to conduct risk segmentation, i.e.\ separating high risk and low risk customers. For this purpose, the expected loss cost in year $T+1$ defined in (\ref{equ:exprate}) is commonly used by the insurer as a risk score to rank and select policyholders. To better illustrate the effects and importance of both temporal and contemporaneous dependence, we conduct analysis for two types of insurance contracts, one without deductibles and one with deductibles.

\textbf{Insurance contracts without deductibles}: Under the case of zero deductibles $d=0$, we have $\eqref{equ:exprate} = \sum_{j=1}^{J} \alpha_{i,T+1}^{(j)} \mathbb{E}(y^{(j)}_{i,T+1}|H_{iT}^{(1)},\ldots,H_{iT}^{(J)}).$ In other words, the expected total loss of the bundled insurance contract can be decomposed into summation of loss cost of each risk. Under such scenario, the temporal dependence within each insurance risk plays an essential role in generating accurate prediction of risk scores of the insurance policy~(i.e.\ expected loss cost in \eqref{equ:exprate}).

We compare two risk scores, one generated by the independence model which ignores dependence among insurance risks~(denoted by $\pi_{ind}$), and one generated  by the proposed D-vine model~(denoted by $\pi_{dep}$). Note that the difference between $\pi_{ind}$ and $\pi_{dep}$ is mainly due to temporal dependence, as the expected total loss cost can be decomposed into summation of expected loss from each peril.



We employ the ordered Lorenz curve (see \cite{FreesMeyersCummings2011}) to assess the out-of-sample performance of the two risk scores in the hold-out sample and demonstrate the importance of temporal dependence within insurance risks. Recall that $\bm{H}_{iT}=(H_{iT}^{(1)},\ldots,H_{iT}^{(J)})$ denotes historical data observed up to period $T$. Let $R_{i,T+1}=R(\bm{H}_{iT})$ be a relativity derived based on $\bm{H}_{iT}$ that the insurer uses to rank individual policyholders, and define $R_{i,T+1}=q_{i,T+1}/p_{i,T+1}$ where $p_{i,T+1}=p(\bm{H}_{iT})$ is a base score that can be interpreted as the premium, and $q_{i,T+1}=q(\bm{H}_{iT})$ is an alternative score that challenges the base. The empirical ordered Lorenz curve describes the relation between $G_L(u)$ and $G_P(u)$ which are defined as:
\begin{align*}
G_L(u) = \frac{\sum_{i=1}^n s_{i,T+1}\mathbb I(R_{i,T+1}\le u)}{\sum_{i=1}^n s_{i,T+1}}~~{\rm and}~~ G_P(u) = \frac{\sum_{i=1}^n p_{i,T+1}\mathbb I(R_{i,T+1}\le u)}{\sum_{i=1}^n p_{i,T+1}}.
\end{align*}
Note that we can interpret $R_{i,T+1}\le u$ as the $i$th policyholder being selected for renewal during underwriting, and thus $G_L$ and $G_P$ can be interpreted as the proportion of losses and premiums of the selected risks at the future period $T+1$.

\begin{figure}[h]
  \begin{center}
   \includegraphics[width=0.38\textwidth,angle=0]{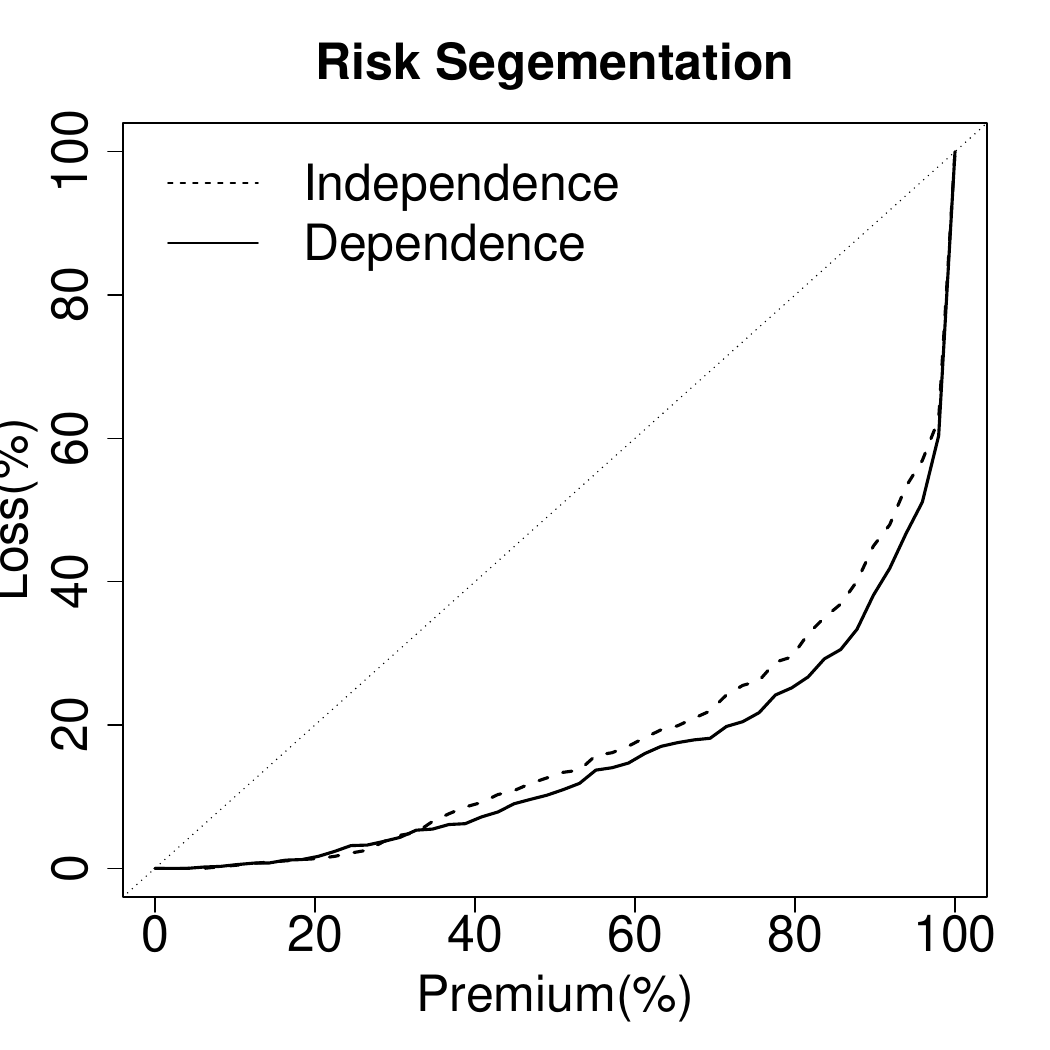}
   \includegraphics[width=0.38\textwidth,angle=0]{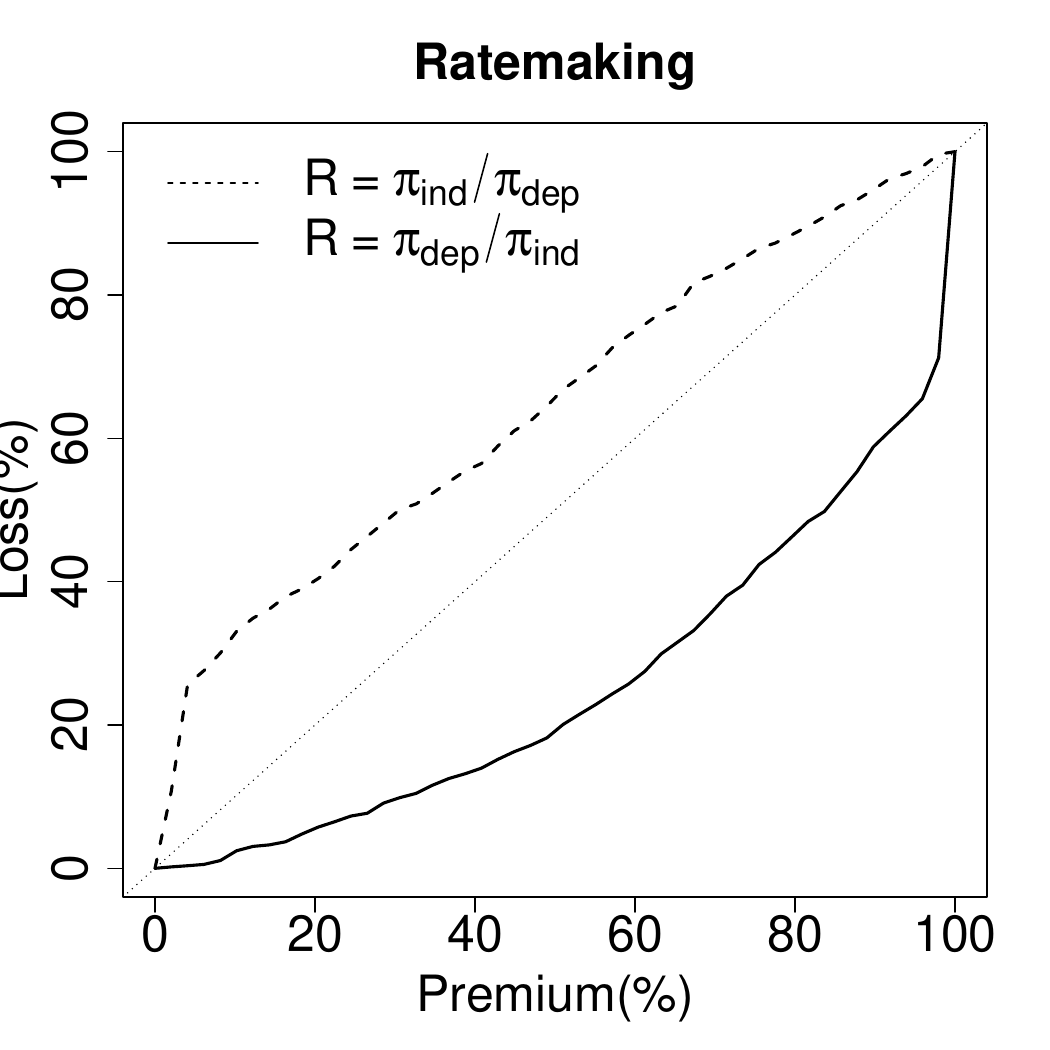}
  \end{center}
  \vspace{-0.8cm}
   \caption{Ordered Lorenz curves for comparing risk scores $\pi_{ind}$ and $\pi_{dep}$ without deductibles. The left panel corresponds to risk segmentation, and the right panel corresponds to ratemaking.}
   \label{fig:gini}
\end{figure}

For the purpose of risk segmentation, we employ the constant premium as the base score, i.e.\ $p_{i,T+1}=1$, and examine which of the two alternative scores $\pi_{ind}$ and $\pi_{dep}$ better identifies profitable portfolio. The corresponding ordered Lorenz curves are shown in the left panel of Figure \ref{fig:gini}. Both curves are below the equity line~(i.e.\ diagonal), indicating substantial opportunities for risk segmentation. For instance, at the 80\% premium level, the proportions of losses of selected contracts are about 30\% and 25\% when using $\pi_{ind}$ and $\pi_{dep}$ for risk segmentation respectively. In addition, there is no crossing between the two ordered Lorenz curves, which suggests that the risk score $\pi_{dep}$ better identifies additional profitable business at every possible underwriting strategy than $\pi_{ind}$. We further summarize the overall performance of the risk score using the associated Gini index calculated as $Gini = 1-2\int_0^{\infty} G_L(u) dG_P(u)$. Graphically speaking, the Gini index is twice the area between the ordered Lorenz curve and the equity line, and thus can be interpreted as average profit over all underwriting strategies. The Gini indices associated with scores $\pi_{ind}$ and $\pi_{dep}$ are 63.97\% and 69.44\% respectively. Thus, appropriate modeling of temporal dependence within each insurance risk provides about 9\% increase in the insurer's profitability.

Furthermore, an insurer can use the ordered Lorenz curve to identify unprofitable business and thus exercises ratemaking to further improve its profitability. In ratemaking, the goal is to set a new insurance premium rate. We show that an insurer can use the risk scores to suggest change to the base premium. For this purpose, we let each of the scores $\pi_{ind}$ and $\pi_{dep}$ serve as the base premium, and the other serve as the challenger. The corresponding Lorenz curves are presented in the right panel of Figure \ref{fig:gini}. In the first case, when $\pi_{ind}$ is the base premium, we observe that the ordered Lorenz curve is below the equity line. This suggests that by switching from the base premium $\pi_{ind}$ to the new premium $\pi_{dep}$, the insurer could better separate high risk and low risk, and thus generate higher profit margin. In contrast, when $\pi_{dep}$ is the base premium, the resulting ordered Lorenz curve lies above the equity line, which corresponds to unprofitable business.

Table \ref{tab:gini} reports the Gini indices associated with Figure \ref{fig:gini}~(right panel). The first row corresponds to the solid line where we use $\pi_{ind}$ as the base premium and use $\pi_{dep}$ as the challenger to create relativity. As confirmed by the positive Gini index, switching from $\pi_{dep}$ to $\pi_{ind}$ for ratemaking results in notable improvement of the insurer's profitability. The second row corresponds to the dashed line where we use $\pi_{dep}$ as the base premium and use $\pi_{ind}$ to create relativity, which hurts the profit as indicated by the negative Gini index. The standard errors in both cases suggest that we can use $\pi_{dep}$ as the new premium rate with confidence.

\begin{table}[h]
  \centering
  \caption{Gini indices for ratemaking without deductibles. Standard errors are in parentheses.}
\begin{tabular}{lrrr}
\hline\hline
      Base &               \multicolumn{3}{c}{  Challenger      }            \\
           & \multicolumn{1}{c}{Independence}            &            & \multicolumn{1}{c}{Dependence}            \\
\cline{2-4}
Independence &                      &            &      49.070     (10.982) \\
Dependence &    -28.317    (13.287) &            &                     \\
\hline\hline
\end{tabular}
  \label{tab:gini}%
\end{table}

{\color{black} To summarize, our analysis shows the importance of modeling temporal dependence within each insurance risk for achieving profitable risk segmentation and pricing in insurance contracts without deductibles. We note that a similar analysis was previously conducted in \cite{ShiYang2018}, where the authors studied temporal dependence for univariate longitudinal data of claim amount and demonstrated its practical value. In comparison, our analysis models temporal dependence for longitudinal data of claim count. Therefore, the two studies complement each other and provide empirical evidence from different angles for the importance of temporal dependence in risk segmentation and pricing.
} 

\textbf{Insurance contracts with deductibles}: An insurance contract often features risk retention measures such as deductible and coverage limit (see e.g.\ \cite{lee2017general}). In particular, deductible is the most common risk retention method that property insurers use to share risks with policyholders. In the following, we consider ratemaking with deductibles and demonstrate the critical role of both temporal and contemporaneous dependence among multivariate longitudinal insurance risks in correctly pricing bundled insurance contracts with deductibles.

We consider an annual deductible $d>0$ which is often found in property insurance and health insurance. Recall from \eqref{equ:exprate} that with a deductible $d$, the expected total loss cost of the insurer for the $i$th policyholder is $\mathbb{E}((S_{i,T+1}-d)_{+}|H_{iT}^{(1)},\ldots,H_{iT}^{(J)})$. Note that the $(\cdot)_+$ operator prevents the decomposition of the total loss into summation of loss cost of each risk. Thus, for an insurance contract with deductibles, we need to calculate its risk score directly from \eqref{equ:exprate}.


We examine two risk scores. The first score is derived from a D-vine based predictive model which only accounts for temporal dependence within each risk~(i.e.\ we set the multivariate copula $C^J$ as an independence copula). This score serves as a proxy for the common practice in the insurance industry for pricing bundled insurance risks. The second score is generated by the proposed D-vine model with both temporal and contemporaneous dependence. A comparison between the two scores highlights the managerial significance of simultaneous analysis of both temporal and contemporaneous dependence on improving decision making of insurers for deductible ratemaking. 

Same as before, we evaluate the out-of-sample ratemaking performance of the two risk scores using the ordered Lorenz curve and the associated Gini index. In particular, we focus on the base-challenger analysis as discussed before, where we use one score as the base premium and use the alternative score as the challenger to create relativity. We test whether switching to the alternative score allows the insurer to better separate low and high risks and improve ratemaking.

\begin{figure}[htp]
	\begin{center}
		\includegraphics[width=0.325\textwidth,angle=0]{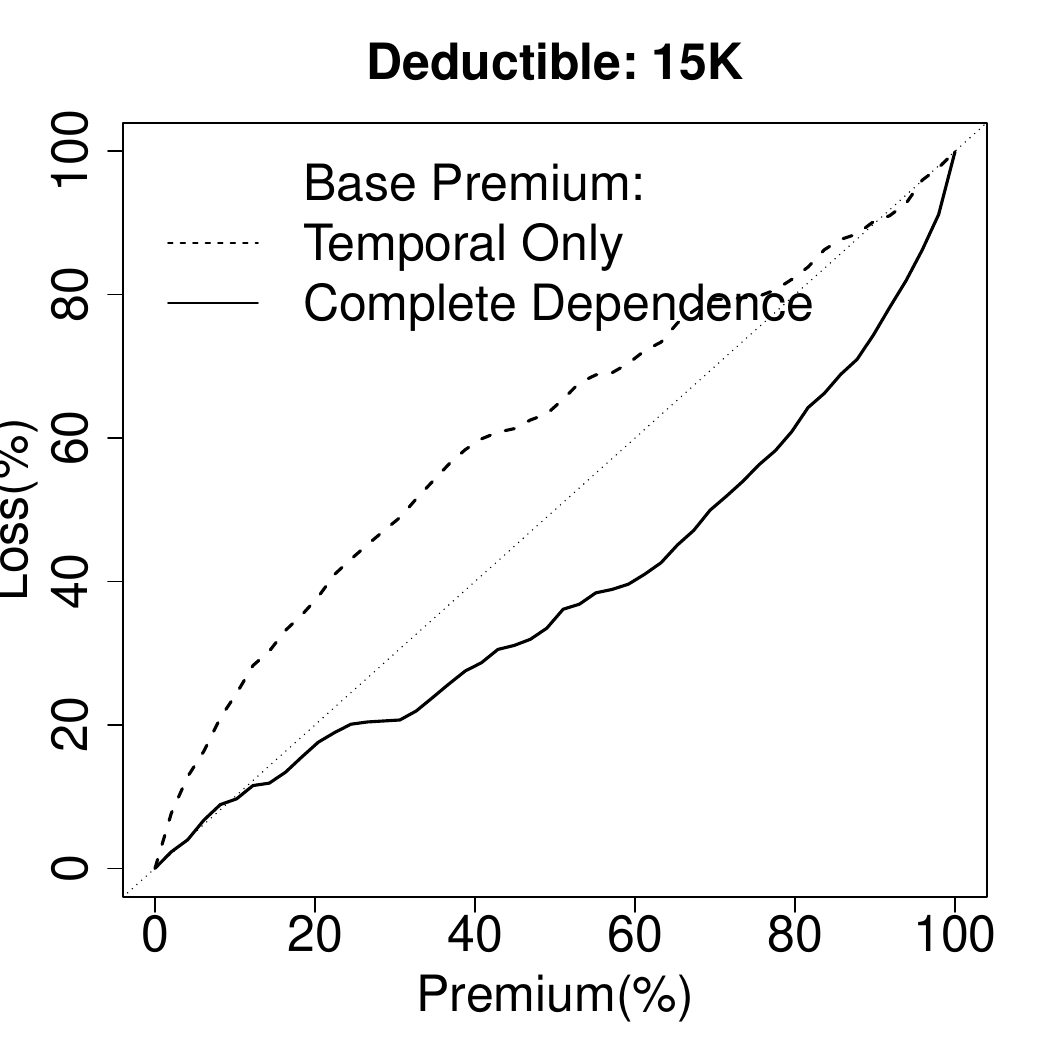}
		\includegraphics[width=0.325\textwidth,angle=0]{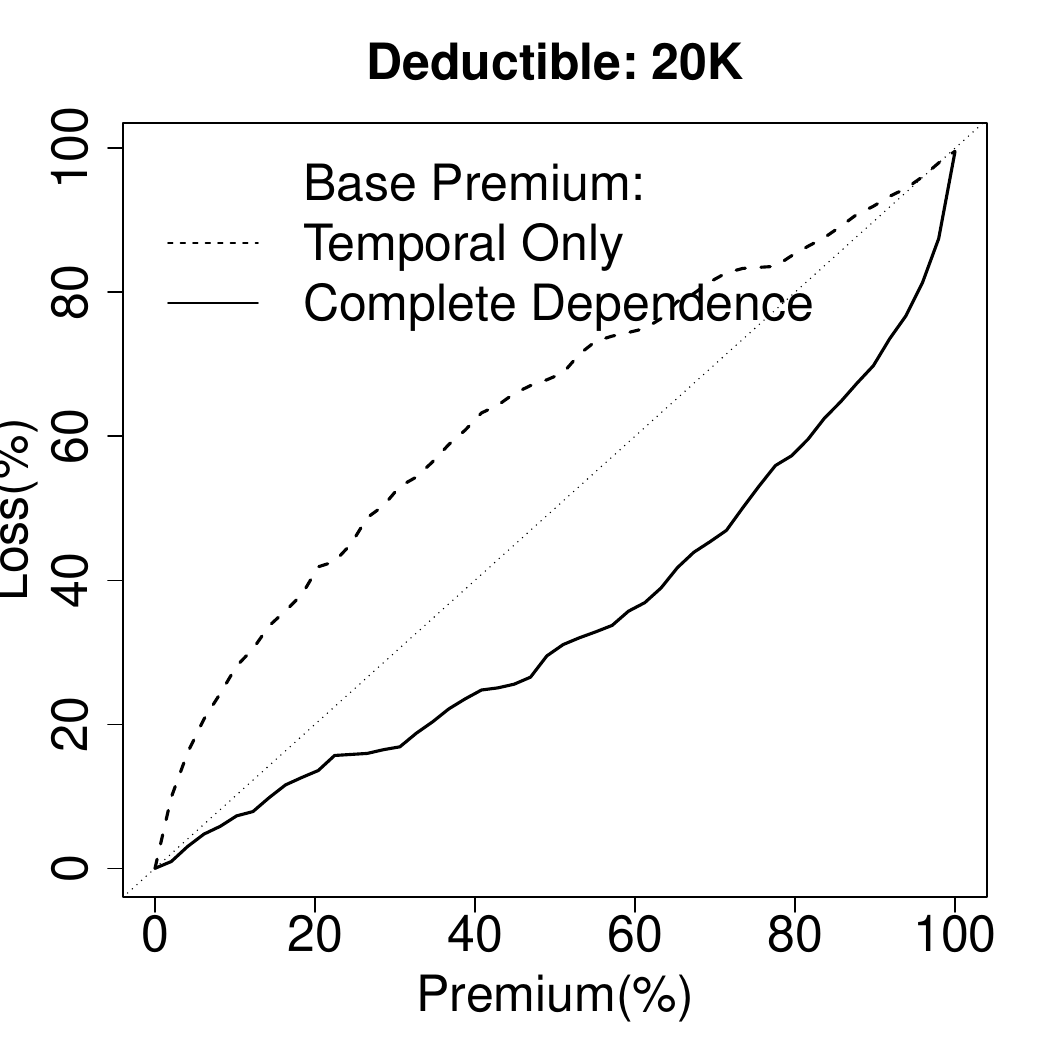}
		\includegraphics[width=0.325\textwidth,angle=0]{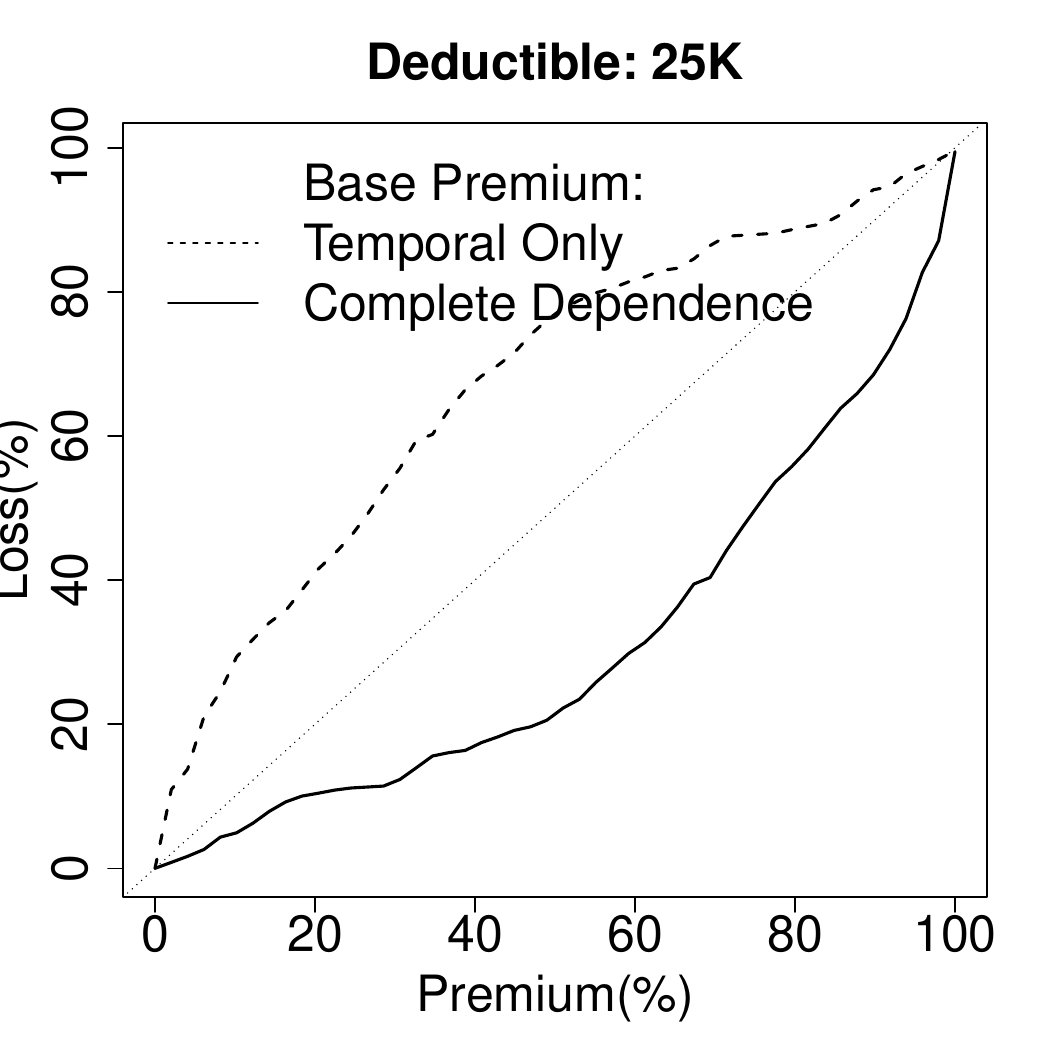}
	\end{center}
	\vspace{-0.8cm}
	\caption{Ordered Lorenz curves for ratemaking at different deductible levels.}
	\label{fig:ginided}
\end{figure}

We consider three different levels of deductibles for the insurance policy with $d$=15K, 20K and 25K. The resulting ordered Lorenz curves and the Gini indices are given in Figure \ref{fig:ginided} and Table \ref{tab:ginideduct} respectively. When the risk score accounting for only temporal dependence is used as the base premium, the Gini indices are positive across all deductible levels, suggesting more profitable ratemaking for the insurer when looking to the alternative score. On the contrary, if the base premium takes into account both temporal and contemporaneous dependence, by switching to the alternative score, the insurer is subject to ratemaking losses as implied by the negative Gini index. In summary, this analysis shows that on top of the temporal dependence, contemporaneous dependence provides additional lift for the insurer to identify profitable business, indicating the importance of simultaneous analysis of both temporal and contemporaneous dependence.

\begin{table}[htbp]
	\centering
	\caption{Gini indices for ratemaking with deductibles. Standard deviations are in parentheses.}
	\begin{tabular}{lrrrrrrrr}
		\hline\hline
		Base &               \multicolumn{8}{c}{  Challenger      }            \\
		&        \multicolumn{3}{c}{Temporal Dep.\ Only}           &            & \multicolumn{3}{c}{Complete Dependence}           &            \\
		&        15K &        20K &        25K &            &        15K &        20K &        25K &            \\
		\cline{2-8}
		Temporal Dep.\ Only &            &            &            &            &     23.140 &     30.530 &     38.060 \\
		&            &            &            &            &      (3.586) &      (3.568) &      (3.632) \\
		Complete Dependence &    -21.910 &    -27.650 &    -34.240 &            &            &            &            \\
		&      (3.605) &      (3.620) &      (3.736) &            &            &            &            \\
		\hline\hline
	\end{tabular}
	\label{tab:ginideduct}%
\end{table}%

\subsection{Risk Management: Portfolio Reinsurance}
In this section, we discuss the application of the D-vine predictive model in portfolio risk management. Risk segmentation and pricing concerns business decisions for individual insurance contracts. In contrast, risk management decisions are made from the viewpoint of an insurance portfolio.

The maximum amount of liability that an insurer could assume determines its underwriting capacity. Quota share reinsurance is a commonly used approach for an insurer to transfer risks to a reinsurance company and thus maintain its underwriting capacity for profitable business. Through reinsurance, the insurer reduces its liability and ensures its ability to pay out claims to policyholders when needed, and thus avoids insolvency.

Under quota share reinsurance, the retained risk by the insurer is given by (\ref{equ:rein}). The insurer's goal is to determine the optimal retention quota $\delta_i$ for each contract $i=1,\cdots,n$ in its insurance portfolio. From the insurer's perspective, the optimal retention $\{\delta_i\}_{i=1}^n$ can be solved via an optimization such that:
\begin{align}
\{\delta_i\}_{i=1}^n = \argmin {\rm Var}(S_{T+1}^{*}|H_1,\ldots,H_T) ~~ s.t.~~ {\mathbb E}(S_{T+1}^{*}|H_1,\ldots,H_T)= K,
\end{align}
where $S_{T+1}^{*} = \sum_{i=1}^n\delta_i S_{i,T+1}$ as defined in \eqref{equ:rein}, $H_t=\{H_{it}^{(j)}: j\in\{1,\ldots,J\}, i\in\{1,\ldots,n\}\}$, and $K$ is a constant representing the target revenue of the insurer and is determined by its underwriting capacity. Essentially, the insurer finds the optimal retention $\{\delta_i\}_{i=1}^n$ by minimizing the volatility of the retained portfolio risk while maintaining a target revenue. Straightforward calculation via the Lagrange multiplier gives that the optimal retention quota for the $i$th contract is
\begin{align}\label{eq:opt_retention}
\delta_i \propto \frac{{\mathbb E}(S_{i,T+1}|\bm{H}_{iT})}{{\rm Var}(S_{i,T+1}|\bm{H}_{iT})} \text{ for } i=1,\cdots,n.
\end{align}
Thus, the optimal quota $\delta_i$ is determined by the conditional distribution of $S_{i,T+1}$, which in turn depends on both temporal and contemporaneous dependence of the bundled insurance risks. 


We compare the performance of two predictive models. The first is a D-vine based predictive model that only accounts for temporal dependence within each risk. This serves as a proxy for the current modeling practice in the insurance industry. The second is the proposed D-vine model which simultaneously accounts for both temporal and contemporaneous dependence.

\begin{figure}[h]
	\begin{center}
		\includegraphics[width=0.38\textwidth,angle=0]{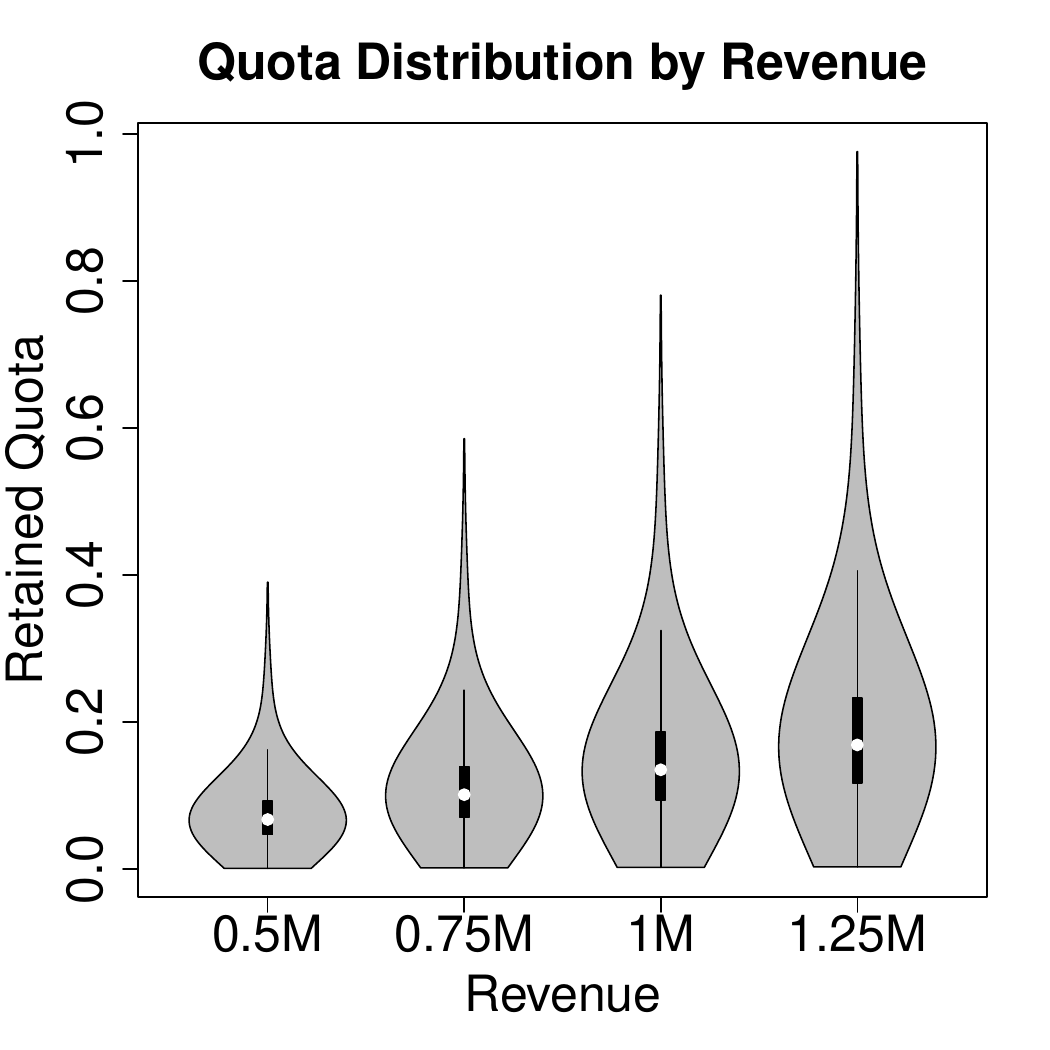}
		\includegraphics[width=0.38\textwidth,angle=0]{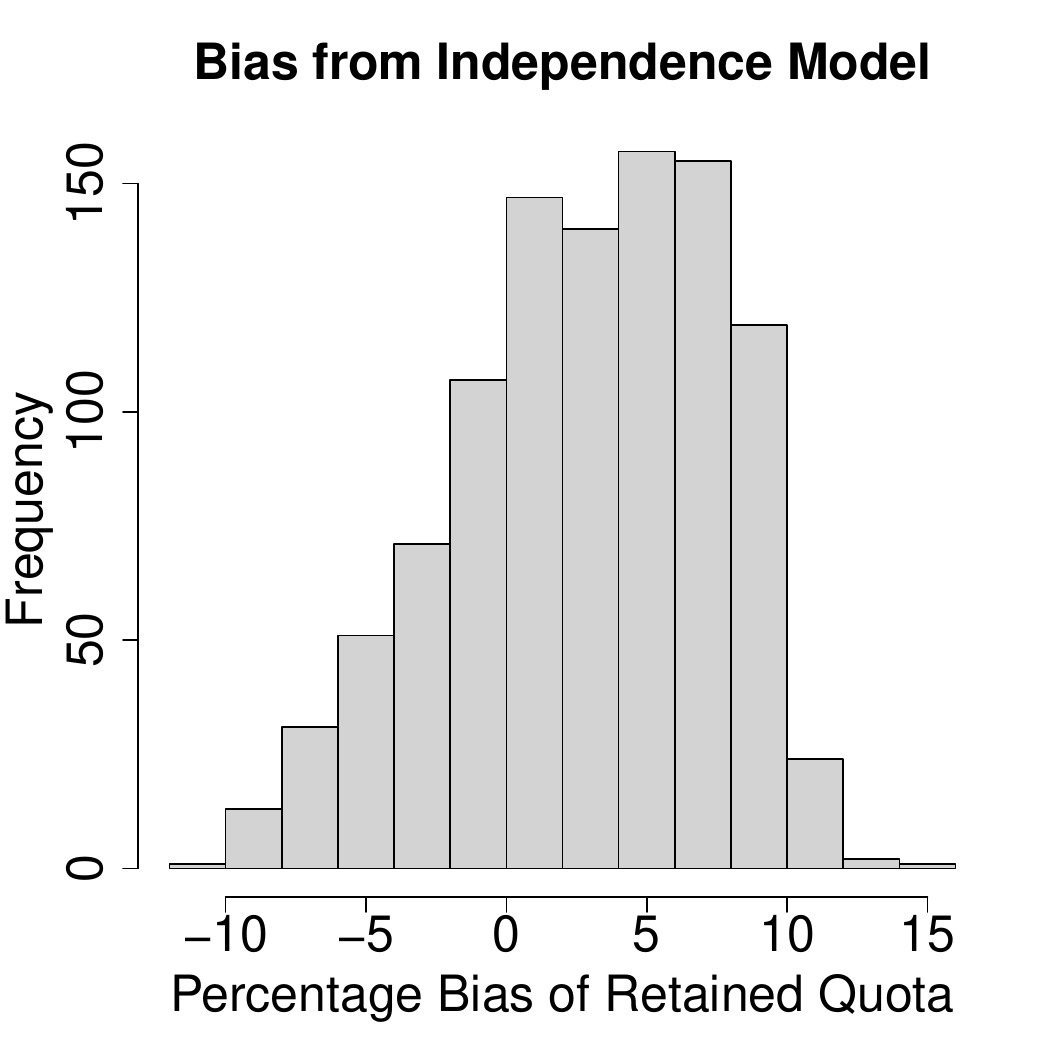}
	\end{center}
	\vspace{-0.8cm}
	\caption{The left panel corresponds to optimal retention quota estimated via the proposed D-vine model. The right panel corresponds to the percentage bias of retention quota caused by the predictive model with only temporal dependence~(i.e.\ ignoring contemporaneous dependence).}
	\label{fig:rein_delta}
\end{figure}

Based on either predictive model, for any fixed target revenue $K$, the insurer can estimate the optimal retention quota $\{\delta_i\}_{i=1}^n$ via \eqref{eq:opt_retention}. In the left panel of Figure \ref{fig:rein_delta}, we show the violin plot of $\{\delta_i\}_{i=1}^n$ estimated via the predicted D-vine model that accounts for both types of dependence at selected levels of target revenue $K$. It is intuitive to observe that larger target revenue $K$ is associated with higher retention quota $\{\delta_i\}_{i=1}^n$ of the insurance risk. Denote $\{\delta_i'\}_{i=1}^n$ as the optimal retention quota estimated via the predictive model with only temporal dependence. To illustrate the impact of contemporaneous dependence, the right panel of Figure \ref{fig:rein_delta} gives the histogram of the percentage bias $\{(\delta_i'-\delta_i)/\delta_i\times 100\%\}_{i=1}^n$ of retained quota estimated from the predictive model with only temporal dependence. As can be seen, ignoring contemporaneous dependence generally results in notable (upward) bias of the estimated retention quota~(see more discussions later). 

\begin{figure}[htp]
  \begin{center}
   \includegraphics[width=0.5\textwidth,angle=0]{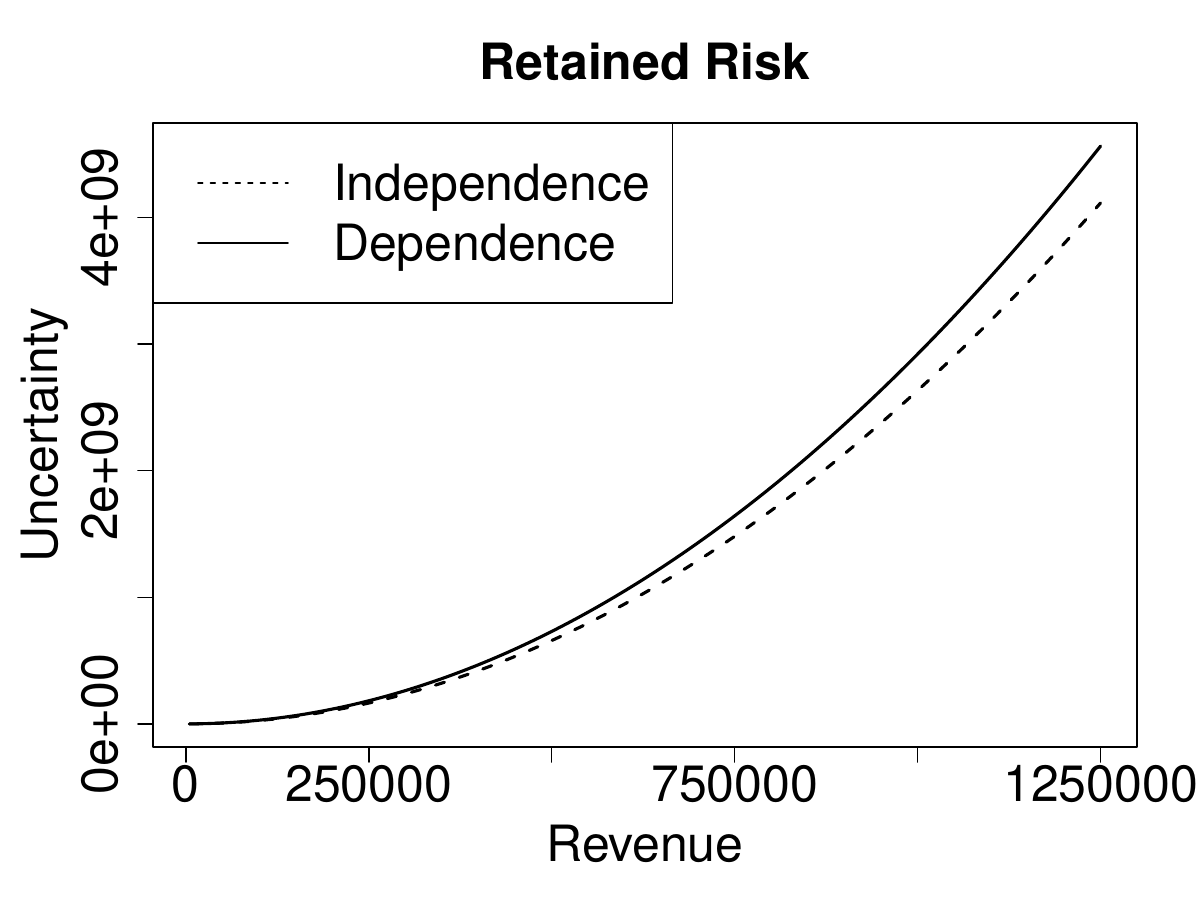}
  \end{center}
  \vspace{-0.8cm}
   \caption{Effects of dependence among bundled insurance risks on risk retention under reinsurance.}
   \label{fig:rein}
\end{figure}

Given $\{\delta_i\}_{i=1}^n$, the insurer can further estimate the volatility of the retained insurance portfolio. The curves in Figure \ref{fig:rein} visualize the relationship between the estimated volatility~(uncertainty) of the optimal retained portfolio and the target revenue $K$ derived based on the two predictive models. As expected, one observes the risk-return trade-off for each curve, i.e.\ with higher underwriting capacity $K$ to assume more business, the insurer undertakes a higher uncertainty in the retained insurance portfolio. 

More importantly, Figure \ref{fig:rein} illustrates the significant effect of the dependence among bundled insurance risks on the reinsurance operation. Note that the estimated volatility of the retained portfolio given by the proposed D-vine model (solid line) is always higher than the volatility estimated by the predictive model with only temporal dependence~(dashed line). In fact, the divergence between the two curves corresponds to a 9.83\% underestimation of the volatility by the latter model. This is indeed not surprising as recall that the multivariate claim counts~(and thus the insurance risks) from the three perils are positively dependent as found in Section \ref{subsec:model_est}~(Table \ref{tab:paircorr}), a phenomenon commonly seen among bundled insurance risks. Thus ignoring the contemporaneous dependence in predictive modeling results in an underestimation of the uncertainty in the retained insurance portfolio, which could lead to disastrous scenarios such as insolvency of the insurer. In summary, this analysis again indicates the managerial significance of simultaneous analysis of both temporal and contemporaneous dependence for insurance operations related to bundled risks.

\vspace{-0.4cm}


\section{Conclusion}\label{sec:conclusion}
In this work, we proposed a D-vine based predictive modeling framework for insurers to manage multivariate insurance risks embedded in insurance policies with bundling features. The proposed framework utilized pair copula construction to allow for simultaneous modeling of the temporal and contemporaneous dependence among multivariate longitudinal insurance risks. Using a dataset on a commercial insurer's portfolio of bundled property insurance policies, we demonstrated the prominent managerial significance of the dependence-aware prediction based on the proposed model. Our work made a methodological contribution to the modeling and analysis of multivariate longitudinal data. Though our analysis focused on the claim count of policyholders, the application of the proposed framework is much broader in that it easily accommodates measurements of various types, be it discrete, semi-continuous, or continuous, and it further allows the multivariate outcomes to be measured in different scales, for instance, a mix of both discrete and continuous outcomes.

In the proposed predictive model, we emphasize two types of dependence among insurance risks, the temporal association within each risk and the contemporaneous dependence across multiple risks. To highlight the managerial implications, we considered two key operations, risk pricing and risk management, that are essential to the insurance business. We showed that dependent risks could have significant impact on the insurer's profitability and on the uncertainty of the insurance portfolio. More importantly, we showed that carefully accounting for the dependence among insurance risks can substantially improve the current practice in the insurance industry. This is of significant practical values, as insurance is an essential sector in any developed economy.

\bibliographystyle{apalike}
\bibliography{ref_MixedVine}

\end{document}


\setlength{\abovedisplayskip}{3pt}
\setlength{\belowdisplayskip}{3pt}
\setlength{\abovedisplayshortskip}{1pt}
\setlength{\belowdisplayshortskip}{1pt}

%
\title{Supplementary Material: Enhanced Pricing and Management of Bundled Insurance Risks with Dependence-aware Prediction using Pair Copula Construction}

\author{
    Peng Shi \\
    Wisconsin School of Business\\
    University of Wisconsin-Madison\\
    \and
    Zifeng Zhao \\
    Mendoza College of Business\\
    University of Notre Dame\\
}
\date{}
\maketitle

\begin{appendices}
	\section{Additional Literature Review}\label{subsec:add_litreview}
	We remark that several strategies have been studied for modeling multivariate longitudinal outcomes in the biostatistics literature, see \cite{Verbeke2014} and \cite{farewell2017two} for recent reviews. However, most methods impose specific structures on the temporal-contemporaneous dependence. In particular, dependence among multivariate longitudinal outcomes is accommodated via one of the three mechanisms.
	
	The first is to use random effects models where latent variables are specified in time-dimension and outcome-dimension to induce dependence. Studies in this line of literature focus on continuous measurements with examples of \cite{Reinsel1984}, \cite{Shah1997}, and \cite{RoyLin2000,RoyLin2002} among others. Conceptually, the random effects model or latent variable model could be extended to non-Gaussian data but is difficult to implement in practice. The second is to directly specify the multivariate distribution for the multidimensional outcomes. The conventional case is the multivariate linear regression for Gaussian outcomes where flexible dependence can be captured by appropriately structuring the covariance matrix (see, for instance, \cite{Galeckii1994}). Specification of the full distribution for discrete outcomes is more challenging, with \cite{MolenberghsLesaffre1994} being one example on ordinal categorical data. The third is the marginal models using generalized estimating equations(GEE)~\citep{LiangZeger1986}. This approach avoids the specification of full distribution and thus inference for regression coefficients is robust with respect to misspecification. For instance, \cite{Rochon1996} considered a bivariate model for a binary outcome and a continuous outcome; \cite{GrayBrookmeyer1998} and \cite{GrayBrookmeyer2000} proposed multidimensional models for continuous, discrete, and time-to-event responses.
	
	However, none of the existing methods can be readily used for modeling multivariate longitudinal insurance risks. First, models with random effects are computationally infeasible for insurance claim counts, and discrete data in general, especially in the setting of nonstandard count regressions. Second, standard multivariate distributions offer limited choices of marginal behavior and dependence structure for the multivariate claim counts. Third, the GEE based marginal model approach treats dependence as nuisance and hence cannot be used for prediction. Setting apart from the current literature, our approach is prediction-oriented, provides flexible modeling for both marginal distribution and dependence structure, and is computationally efficient for discrete observations.

	\section{Theoretical Guarantees}\label{sec:thm}
	In this section, we establish theoretical guarantees for the asymptotic consistency and normality of $\widehat{\bftheta}=(\widehat{\bfbeta},\widehat{\bfzeta},\widehat{\bfrho})$. Same as the standard longitudinal literature, our theoretical result is established under the asymptotic setting of fixed $T$ and diverging $n$. Before stating the theoretical result, for the ease of presentation, we first introduce some more notations in addition to the ones defined in Section \ref{sec:inference} of the main text.
	
	Denote the true model parameter as $\bftheta^o=(\bfbeta^o,\bfzeta^o,\bfrho^o)$ and denote the parameter space of $\bftheta$ as $\bm{\Theta}.$ For the first stage estimator $\widehat{\bfbeta}=(\widehat{\bfbeta}_1,\cdots,\widehat{\bfbeta}_J)$, define for $j=1,\cdots, J$,
	$$B_1^{(j)}=-\mathbb{E}\left(\frac{\partial^2}{\partial \bfbeta_j \partial \bfbeta_j^\top}l_{1i}^{(j)}(\bfbeta_j^o)\right) \text{ and } \Sigma_1^{(j)}=\text{Var}\left(\frac{\partial}{\partial \bfbeta_j}l_{1i}^{(j)}(\bfbeta_j^o)\right).$$
	For the second stage estimator $\widehat{\bfzeta}=(\widehat{\bfzeta}_1.\cdots,\widehat{\bfzeta}_J)$, define for $j=1,\cdots, J$,
	\begin{align*}
	&B_2^{(j)}=-\mathbb{E}\left(\frac{\partial^2}{\partial \bfzeta_j \partial \bfzeta_j^\top}l_{2i}^{(j)}(\bfzeta_j^o;\bfbeta_j^o)\right), D_2^{(j)}=\mathbb{E}\left(\frac{\partial^2}{\partial \bfzeta_j \partial \bfbeta_j^\top}l_{2i}^{(j)}(\bfzeta_j^o;\bfbeta_j^o)\right),\\
	&\Sigma_2^{(j)}=\text{Var}\left(\frac{\partial}{\partial \bfzeta_j}l_{2i}^{(j)}(\bfzeta_j^o;\bfbeta_j^o)+D_2^{(j)}(B_1^{(j)})^{-1} \frac{\partial}{\partial \bfbeta_j}l_{1i}^{(j)}(\bfbeta_j^o)\right).
	\end{align*}
	For the third stage estimator $\widehat{\bfrho}$, define
	\begin{align*}
		&B_3=-\mathbb{E}\left(\frac{\partial^2}{\partial \bfrho \partial \bfrho^\top}l_{i}(\bfbeta^o,\bfzeta^o,\bfrho^o)\right), D_{31}^{(j)}=\mathbb{E}\left(\frac{\partial^2}{\partial \bfrho \partial \bfbeta_j^\top}l_{i}(\bfbeta^o,\bfzeta^o,\bfrho^o)\right),
		D_{32}^{(j)}=\mathbb{E}\left(\frac{\partial^2}{\partial \bfrho \partial \bfzeta_j^\top}l_{i}(\bfbeta^o,\bfzeta^o,\bfrho^o)\right),\\
	    &\Sigma_3=\text{Var}\left(\frac{\partial}{\partial \bfrho}l_{i}(\bfbeta^o,\bfzeta^o,\bfrho^o)+\sum_{j=1}^J D_{31}^{(j)}(B_1^{(j)})^{-1} \frac{\partial}{\partial \bfbeta_j}l_{1i}^{(j)}(\bfbeta_j^o)+\sum_{j=1}^J
	    D_{32}^{(j)}(B_2^{(j)})^{-1} \frac{\partial}{\partial \bfzeta_j}l_{2i}^{(j)}(\bfzeta_j^o;\bfbeta_j^o)\right).
	\end{align*}	
	
	We further introduce some mild regularity conditions on the parameter space $\bf \Theta$ and the likelihood functions, which are standard assumptions for likelihood-based estimators.
	
	\noindent\textbf{Assumption 1:}
	\vspace{-0.3cm}
	\begin{itemize}
		\item A1. The $n$ multivariate longitudinal observations are independently and identically generated from the true model with parameter $\bftheta^o$.
		\item A2. The parameter space $\bf\Theta$ is a compact set and the true parameter $\bftheta^o\in \bf\Theta$.
		\item A3. The likelihood functions $\{l_{1i}^{(j)}(\bfbeta_j)\}_{j=1}^J$, $\{l_{2i}^{(j)}(\bfzeta_j;\bfbeta_j)\}_{j=1}^J$ and $l_{i}(\bfbeta,\bfzeta,\bfrho)$ are twice continuously differentiable w.r.t. its arguments for all $i=1,\cdots,n.$
		\item A4. (a). The population likelihood function $\{\mathbb{E}(l_{1i}^{(j)}(\bfbeta_j))\}_{j=1}^J$, $\{\mathbb{E}(l_{2i}^{(j)}(\bfzeta_j;\bfbeta_j))\}_{j=1}^J$ and $\mathbb{E}(l_{i}(\bfbeta,\bfzeta,\bfrho))$ are uniquely maximized at the true parameter $\{\bfbeta_j^o\}_{j=1}^J$, $\{\bfbeta_j^o,\bfzeta_j^o\}_{j=1}^J$ and $\bftheta^o$ respectively. (b). In addition, we have for $j=1,\cdots,J,$
		$$\mathbb{E}\left(\sup_{\bfbeta_j}\left|l_{1i}^{(j)}(\bfbeta_j)\right|\right),\mathbb{E}\left(\sup_{\bfbeta_j,\bfzeta_j}\left|l_{2i}^{(j)}(\bfzeta_j;\bfbeta_j)\right|\right),\mathbb{E}\left(\sup_{\bfbeta,\bfzeta,\bfrho}\left|l_i(\bfbeta,\bfzeta,\bfrho)\right|\right)<\infty. $$
		\item A5. The covariance matrices $\Sigma_1^{(j)}$, $\Sigma_2^{(j)}$ and $\Sigma_3$ are well-defined for $j=1,2,\cdots,J.$
		\item A6. The second-order derivatives are bounded in expectation where for $j=1,\cdots,J,$ we have
		$$\mathbb{E}\left(\sup_{\bfbeta_j}\left\|\frac{\partial^2}{\partial \bfbeta_j \partial \bfbeta_j^\top}l_{1i}^{(j)}(\bfbeta_j)\right\|\right),\mathbb{E}\left(\sup_{\bfbeta_j,\bfzeta_j}\left\|\frac{\partial^2}{\partial \bfzeta_j \partial \bfzeta_j^\top}l_{2i}^{(j)}(\bfzeta_j;\bfbeta_j)\right\|\right), \mathbb{E}\left(\sup_{\bfbeta_j,\bfzeta_j}\left\|\frac{\partial^2}{\partial \bfzeta_j \partial \bfbeta_j^\top}l_{2i}^{(j)}(\bfzeta_j;\bfbeta_j)\right\|\right)<\infty,$$
		$$\mathbb{E}\left(\sup_{\bfrho}\left\|\frac{\partial^2}{\partial \bfrho \partial \bfrho^\top}l_{i}(\bfbeta^o,\bfzeta^o,\bfrho)\right\|\right),\mathbb{E}\left(\sup_{\bfbeta,\bfrho}\left\|\frac{\partial^2}{\partial \bfrho \partial \bfbeta^\top}l_{i}(\bfbeta,\bfzeta^o,\bfrho)\right\|\right), \mathbb{E}\left(\sup_{\bfzeta,\bfrho}\left\|\frac{\partial^2}{\partial \bfrho \partial \bfzeta^\top}l_{i}(\bfbeta^o,\bfzeta,\bfrho)\right\|\right)<\infty.$$
	\end{itemize}
    Assumptions A1-A6 are commonly seen in the literature of likelihood-based estimation. In particular, A4 and A6 are used to invoke the uniform law of large numbers~(ULLN) in the proof.
	
	\begin{theorem}\label{thm:mle}
		(i) Under assumptions A1-A6, we have $\|\widehat{\bftheta} -\bftheta^o\|=o_p(1), $ i.e.  $\hat{\bftheta}$ is consistent.
		
		\noindent (ii) Under assumptions A1-A6 and in addition that $\bftheta^o\in$ interior$(\bm{\Theta})$, we have: (1) $\sqrt{n}(\widehat{\bfbeta}_j -\bfbeta_j^o) \to N(0, (B_1^{(j)})^{-1}\Sigma_1^{(j)} (B_1^{(j)})^{-1})$ in distribution for $j=1,\cdots,J$. (2) $\sqrt{n}(\widehat{\bfzeta}_j -\bfzeta_j^o) \to N(0, (B_2^{(j)})^{-1}\Sigma_2^{(j)} (B_2^{(j)})^{-1})$ in distribution for $j=1,\cdots,J$. (3)  $\sqrt{n}(\widehat{\bfrho} -\bfrho^o) \to N(0, B_3^{-1}\Sigma_3 B_3^{-1})$ in distribution.
	\end{theorem}

    Several remarks regarding Theorem \ref{thm:mle} are in order. First, same as the classical MLE, the three-stage MLE $\widehat{\bftheta}$ is consistent and admits an asymptotic error of order $O_p(1/\sqrt{n})$. Second, different from the classical MLE, the asymptotic covariance matrix of $\widehat{\bftheta}$ is much more complicated due to the (incorrect) working independence assumptions used in its estimation for the purpose of computational efficiency. In particular, all three-stage estimators have a sandwich-type asymptotic covariance, which is commonly seen under model misspecification~\citep{Godambe1960}. Third, as can be seen from Theorem \ref{thm:mle}(ii), the estimation error of $\widehat{\bfbeta}$ enters the asymptotic covariance of $\widehat{\bfzeta}$, and similarly the estimation error of $\widehat{\bfbeta}$ and $\widehat{\bfzeta}$ affects the asymptotic covariance of $\widehat{\bfrho}$. Intuitively, this is due to the sequential estimation procedure, as we fix $\widehat{\bfbeta}$ in the estimation of $\bfzeta$, and fix $\widehat{\bfbeta}$, $\widehat{\bfzeta}$ in the estimation of $\bfrho$.

    Indeed, the three-stage MLE can be viewed under the framework of the so-called \textit{two-step} estimator in the econometrics literature, where the estimation of model parameters is conducted in a step-wise fashion. We refer to \cite{Newey1994} for a general treatment of the \textit{two-step} estimator, where similar phenomenon and asymptotic results as that in Theorem \ref{thm:mle} are discussed. \\

    \noindent\textbf{Proof of Theorem \ref{thm:mle}}: We start with the first stage estimator $\widehat{\bfbeta}_j$ for $j=1,\cdots,J$. The proof for $\widehat{\bfbeta}_j$ follows the standard technique for the classical maximum likelihood estimation. For the consistency result, by A1 and A4(b), we can invoke the uniform law of large numbers~(ULLN) for the likelihood function such that
    $$\sup_{\bfbeta_j}\left|\frac{1}{n}L_1^{(j)}(\bfbeta_j) - \mathbb{E}(l_{1i}^{(j)}(\bfbeta_j)) \right|\to 0$$
    in probability. Combined with the compactness of the parameter space $\bf \Theta$ in A2, this implies that the maximizer of $\frac{1}{n}L_1^{(j)}(\bfbeta_j)$, which is $\widehat{\bfbeta}_j$, converges to the maximizer of $\mathbb{E}(l_{1i}^{(j)}(\bfbeta_j))$, which is $\bfbeta_j^o$ as stated in A4(a). Thus, we have $\widehat{\bfbeta}_j\to_p \bfbeta_j^o$ for $j=1,\cdots, J.$

    For the asymptotic normality, by the definition of $\widehat{\bfbeta}_j$ and Taylor expansion, we have
    \begin{align*}
    	0=\frac{1}{\sqrt{n}}\frac{\partial }{\partial \bfbeta_j} L_1^{(j)}(\widehat{\bfbeta}_j)=\frac{1}{\sqrt{n}}\frac{\partial }{\partial \bfbeta_j} L_1^{(j)}({\bfbeta}_j^o)+\frac{1}{n}\frac{\partial^2}{\partial \bfbeta_j \partial \bfbeta_j^\top } L_1^{(j)}(\widetilde{\bfbeta}_j)\sqrt{n}(\widehat{\bfbeta}_j-\bfbeta_j^o),
    \end{align*}
    where $\widetilde{\bfbeta}_j$ is some value between $\widehat{\bfbeta}_j$ and $\bfbeta_j^o$. By the central limit theorem and A5, we have that
    $$\frac{1}{\sqrt{n}}\frac{\partial }{\partial \bfbeta_j} L_1^{(j)}({\bfbeta}_j^o)=\frac{1}{\sqrt{n}} \sum_{i=1}^{n}\frac{\partial }{\partial \bfbeta_j} l_{1i}^{(j)}(\bfbeta_j^o)\to N(0,\Sigma_1^{(j)})$$
    in distribution. Furthermore by the consistency of $\widehat{\bfbeta}_j$ and A6, we can invoke ULLN for the second order derivative and obtain
    $$-\frac{1}{n}\frac{\partial^2}{\partial \bfbeta_j \partial \bfbeta_j^\top } L_1^{(j)}(\widetilde{\bfbeta}_j) \to B_1^{(j)}$$
    in probability. Thus, simple algebra gives
    $$\sqrt{n}(\widehat{\bfbeta}_j -\bfbeta_j^o) \to N(0, (B_1^{(j)})^{-1}\Sigma_1^{(j)} (B_1^{(j)})^{-1})$$ in distribution for $j=1,\cdots,J.$

    We now turn to the proof of the second stage estimator $\widehat{\bfzeta}_j$ for $j=1,\cdots,J$. The proof for consistency is similar as before. Specifically, by A1 and A4(b) and the consistency of $\widehat{\bfbeta}_j$, we can invoke the ULLN and obtain
    $$\sup_{\bfzeta_j}\left|\frac{1}{n}L_2^{(j)}(\bfzeta_j;\widehat{\bfbeta}_j) - \mathbb{E}(l_{2i}^{(j)}(\bfzeta_j;{\bfbeta}_j^o)) \right|\to 0$$
    in probability, which implies that the maximizer of $\frac{1}{n}L_2^{(j)}(\bfzeta_j;\widehat{\bfbeta}_j)$, which is $\widehat{\bfzeta}_j$, converges to the maximizer of $\mathbb{E}(l_{2i}^{(j)}(\bfzeta_j;{\bfbeta}_j^o))$, which is $\bfzeta_j^o$ as stated in A4(a). Thus, we have $\widehat{\bfzeta}_j\to_p \bfzeta_j^o$ for $j=1,\cdots, J.$

    The proof for asymptotic normality is more complicated as we need to further control the estimation error due to the first stage estimator $\widehat{\bfbeta}_j$ by an extra Taylor expansion. Specifically, by the definition of $\widehat{\bfbeta}_j$ and two Taylor expansion, we have
        \begin{align*}
    	0&=\frac{1}{\sqrt{n}}\frac{\partial }{\partial \bfzeta_j} L_2^{(j)}(\widehat{\bfzeta}_j; \widehat{\bfbeta}_j)=\frac{1}{\sqrt{n}}\frac{\partial }{\partial \bfzeta_j} L_2^{(j)}({\bfzeta}_j^o; \widehat{\bfbeta}_j)+\frac{1}{n}\frac{\partial^2}{\partial \bfzeta_j \partial \bfzeta_j^\top } L_2^{(j)}(\widetilde{\bfzeta}_j;\widehat{\bfbeta}_j)\sqrt{n}(\widehat{\bfzeta}_j-\bfzeta_j^o)\\
    	&=\frac{1}{\sqrt{n}}\frac{\partial }{\partial \bfzeta_j} L_2^{(j)}({\bfzeta}_j^o; {\bfbeta}_j^o)+ \frac{1}{{n}}\frac{\partial^2 }{\partial \bfzeta_j\partial\bfbeta_j^\top} L_2^{(j)}({\bfzeta}_j^o; \widetilde{\bfbeta}_j)\sqrt{n}(\widehat{\bfbeta}_j-\bfbeta_j^o)+  \frac{1}{n}\frac{\partial^2}{\partial \bfzeta_j \partial \bfzeta_j^\top } L_2^{(j)}(\widetilde{\bfzeta}_j;\widehat{\bfbeta}_j)\sqrt{n}(\widehat{\bfzeta}_j-\bfzeta_j^o),
    \end{align*}
    where $\widetilde{\bfbeta}_j$ is a value between $\widehat{\bfbeta}_j$ and $\bfbeta_j^o$ and $\widetilde{\bfzeta}_j$ is a value between $\widehat{\bfzeta}_j$ and $\bfzeta_j^o$. By the central limit theorem and A5, combined with the asymptotic normality of $\widehat{\bfbeta}_j$ established earlier, we have that
    \begin{align*}
    	&\frac{1}{\sqrt{n}}\frac{\partial }{\partial \bfzeta_j} L_2^{(j)}({\bfzeta}_j^o; {\bfbeta}_j^o)+\frac{1}{{n}}\frac{\partial^2 }{\partial \bfzeta_j\partial\bfbeta_j} L_2^{(j)}({\bfzeta}_j^o; \widetilde{\bfbeta}_j)\sqrt{n}(\widehat{\bfbeta}_j-\bfbeta_j^o)\\
    	=&\frac{1}{\sqrt{n}} \sum_{i=1}^{n} \left(\frac{\partial }{\partial \bfzeta_j}l_{2i}^{(j)}(\bfzeta_j^o;\bfbeta_j^o) +D_2^{(j)}(B_1^{(j)})^{-1} \frac{\partial}{\partial \bfbeta_j}l_{1i}^{(j)}(\bfbeta_j^o)\right)+o_p(1)\to N(0,\Sigma_2^{(j)})
    \end{align*}
    in distribution. Furthermore by the consistency of $\widehat{\bfzeta}_j$ and A6, we can invoke ULLN for the second order derivative and obtain
    $$\frac{1}{n}\frac{\partial^2}{\partial \bfzeta_j \partial \bfzeta_j^\top } L_2^{(j)}(\widetilde{\bfzeta}_j;\widehat{\bfbeta}_j) \to B_2^{(j)}$$
    in probability. Thus, simple algebra gives
    $\sqrt{n}(\widehat{\bfzeta}_j -\bfzeta_j^o) \to N(0, (B_2^{(j)})^{-1}\Sigma_2^{(j)} (B_2^{(j)})^{-1})$ in distribution for $j=1,\cdots,J$.

    The proof for the third stage estimator $\widehat{\bfrho}$ follows the same arguments as the one for the second stage estimator $\widehat{\bfzeta}_j$ but requires an additional Taylor expansion as both first and second-stage estimator will contribute to the estimation error of $\widehat{\bfrho}$. We skip the proof as the extension is straightforward.

\section{Numerical Experiments}\label{sec:simulation}
This section examines the performance of the statistical estimation and inference procedures proposed in Section \ref{sec:inference} of the main text via numerical experiments. Specifically, Section \ref{subsec:simu_paraest} investigates the finite-sample performance of the three-stage MLE and Section \ref{subsec:simu_copselection} investigates the performance of the tree-by-tree sequential D-vine selection procedure.

\subsection{Parameter Estimation}\label{subsec:simu_paraest}
In this section, we examine the performance of the three-stage MLE and the asymptotic covariance estimation based on parametric bootstrap. The multivariate longitudinal data is generated according to the D-vine predictive model proposed in Section \ref{subsec:riskmodel} of the main text. Specifically, for each policyholder $i=1,\cdots,n$, we assume that there are three bundled risks ($J=3$), which correspond to the three insurance risks (water, fire, and others) in our empirical study, and each policyholder is observed for four years ($T=4$). The three components of the model are specified as below:

\begin{enumerate}[{(1)}]
	\item (Marginal regression) The marginal count regression model for $y_{it}^{(j)}$ follows a Poisson generalized linear model~(a special case of the negative binomial regression) with intensity
	$$\log \left(\lambda_{ijt}\right) = \beta_{j0} + \beta_{j1}x_{1,it}^{(j)} + \beta_{j2}x_{2,i}^{(j)}.$$
	We set $\bfbeta_j=(\beta_{j0},\beta_{j1},\beta_{j2})=(-1,0.5,0.5)$ for $j=1,2,3$, and set $x_{1,it}^{(j)}\overset{i.i.d.}{\sim}N(0,1)$ and $x_{2,ij}\overset{i.i.d.}{\sim}Bernoulli(0.4)$ to generate a continuous and a discrete exogenous predictor.
	\item (D-vine) For $j=1$, the bivariate copulas of the D-vine are set to be rotated Gumbel copula for tree 1 (Kendall's $\tau=0.5$), rotated Joe copula for tree 2 (Kendall's $\tau=0.3$), and rotated Joe copula for tree 3 (Kendall's $\tau=0.1$), which gives $\bfzeta_1=(\zeta_{11}, \zeta_{12}, \zeta_{13})=(2, 1.77, 1.19)$ for model parameters of the bivariate copulas in the three trees\footnote{Note that for all bivariate copulas considered in the simulation study, there is a one-to-one relationship between its model parameter and its Kendall's $\tau$. We use Kendall's $\tau$ in model specification for better clarity.}. For $j=2$, we set rotated Gumbel for tree 1 (Kendall's $\tau=0.6$), rotated Gumbel for tree 2 (Kendall's $\tau=0.4$), and rotated Clayton for tree 3 (Kendall's $\tau=0.2$), which gives $\bfzeta_2=(\zeta_{21}, \zeta_{22}, \zeta_{23})=(2.5, 1.67, 0.5)$. For $j=3$, we set rotated Gumbel for tree 1 (Kendall's $\tau=0.7$), rotated Clayton for tree 2 (Kendall's $\tau=0.5$), and rotated Joe for tree 3 (Kendall's $\tau=0.3$), which gives $\bfzeta_3=(\zeta_{31}, \zeta_{32}, \zeta_{33})=(3.33, 2, 1.77)$.
	\item (Multivariate copula $C_J$) The contemporaneous copula $C_J$ is set to be a Gaussian copula with unstructured dispersion matrix with $\bfrho=(\rho_{12},\rho_{13},\rho_{23})=(0.2,0.5,0.8)$, where $\rho_{jj'}$ denotes the pair-wise correlation between the claim counts from the $j$th and $j'$th insurance risks.
\end{enumerate}


We conduct simulation experiments with sample size $n$ (number of policyholders) to be 500 and 1000. For each level of $n$, we repeat the experiment 500 times. The three-stage MLE $\widehat{\bftheta}$ is used for parameter estimation and we further construct confidence intervals~(C.I.) for the true model parameter using the estimated asymptotic covariance based on parametric bootstrap. The results are summarized in Table \ref{MLECount}, which reports the sample mean and sample standard deviation of $\widehat{\bftheta}$ across the 500 experiments, as well as the sample coverage probabilities of the constructed C.I.\ at different confidence levels.

To conserve space, we only report the estimation results for the parameters $\bfzeta$ in the D-vine and $\bfrho$ in the multivariate copula $C_J$, which are more challenging to estimate than $\bfbeta$ in the marginal count regression. As can be seen clearly, the three-stage MLE is consistent and achieves satisfactory performance for $n=500$. In addition, both the estimation accuracy and the coverage probability of the C.I.\ are improving with the increase of sample size $n$.


\begin{table}[h]
	\centering
	\caption{Performance of the three-stage MLE. Mean stands for the sample mean of the MLE across 500 experiments. S.D.\ stands for the sample standard deviation of the MLE across 500 experiments. CI gives the sample coverage rate of the confidence interval constructed based on parametric bootstrap across 500 experiments.}
	\begin{tabular}{ccccccccccccc}\hline\hline
		$n=500$    & $\zeta_{11}$ & $\zeta_{12}$ & $\zeta_{13}$ & $\zeta_{21}$ & $\zeta_{22}$ & $\zeta_{23}$ & $\zeta_{31}$ & $\zeta_{32}$ &  $\zeta_{33}$ & $\rho_{12}$  & $\rho_{13}$ & $\rho_{23}$    \\\hline
		$\text{Mean}$ & 2.01 & 1.77 & 1.19 & 2.50 & 1.67 & 0.50 & 3.35 & 1.99 & 1.82 & 0.20 & 0.50 & 0.80\\
		$\text{S.D.}$ & 0.09 & 0.13 & 0.12 & 0.13 & 0.08 & 0.11 & 0.19 & 0.21 & 0.26 & 0.04 & 0.04 & 0.02 \\
		$\text{CI 90\%}$ & 0.90 & 0.88 & 0.84 & 0.89 & 0.89 & 0.88 & 0.93 & 0.89 & 0.92 & 0.87 & 0.89 & 0.89\\
		$\text{CI 95\%}$ & 0.95 & 0.94 & 0.90 & 0.93 & 0.96 & 0.95 & 0.97 & 0.94 & 0.96 & 0.91 & 0.94 & 0.95 \\
		$\text{CI 99\%}$ & 0.99 & 0.98 & 0.96 & 0.97 & 0.99 & 0.98 & 0.99 & 0.99 & 0.99 & 0.96 & 0.98 & 0.99 \\\hline
		$n=1000$    & $\zeta_{11}$ & $\zeta_{12}$ & $\zeta_{13}$ & $\zeta_{21}$ & $\zeta_{22}$ & $\zeta_{23}$ & $\zeta_{31}$ & $\zeta_{32}$ &  $\zeta_{33}$ & $\rho_{12}$  & $\rho_{13}$ & $\rho_{23}$    \\\hline
		$\text{Mean}$ & 2.00 & 1.76 & 1.19 & 2.50 & 1.67 & 0.50 & 3.34 & 1.99 & 1.78 & 0.20 & 0.50 & 0.80 \\
		$\text{S.D.}$ & 0.06 & 0.09 & 0.08 & 0.09 & 0.05 & 0.08 & 0.13 & 0.15 & 0.16 & 0.03 & 0.02 & 0.02 \\
		$\text{CI 90\%}$ & 0.88 & 0.90 & 0.89 & 0.90 & 0.91 & 0.88 & 0.92 & 0.88 & 0.91 & 0.92 & 0.92 & 0.92 \\
		$\text{CI 95\%}$ & 0.92 & 0.95 & 0.93 & 0.95 & 0.94 & 0.94 & 0.97 & 0.94 & 0.95 & 0.95 & 0.95 & 0.95 \\
		$\text{CI 99\%}$ & 0.98 & 0.98 & 0.98 & 0.99 & 0.98 & 0.99 & 0.99 & 0.98 & 0.99 & 0.98 & 0.99 & 0.99 \\ \hline\hline
	\end{tabular}
	\label{MLECount}
\end{table}

\subsection{Sequential Copula Selection in D-vine}\label{subsec:simu_copselection}
In this section, we investigate the performance of the tree-by-tree sequential selection procedure for bivariate copulas  $\{\{C_{s,t}^{(j)}\}_{s=1}^{t-1}\}_{t=2}^T$ of the D-vine. We conduct numerical experiments for $J=3$ D-vines with different parameter settings. The data generating process is almost the same as the one in Section \ref{subsec:simu_paraest}. We keep the marginal count regression~(component (1)) and the multivariate copula $C_J$~(component (3)) the same and the only difference is the D-vine~(component (2)).

In particular, to examine the performance of the sequential selection procedure in terms of correctly determining the optimal credibility weight to the historical measurements~(i.e. the truncation mechanism discussed in Section \ref{subsec:inference} of the main text), we set the length of longitudinal observations to be $T=5$ and we set the three D-vines to be truncated at different tree levels.

Specifically, for $j=1$, the D-vine is truncated at tree 2, where the bivariate copulas are set to be rotated Gumbel for tree 1 (Kendall's $\tau=0.5$) and rotated Joe for tree 2 (Kendall's $\tau=0.3$). For $j=2$, the D-vine is truncated at tree 3, we set rotated Gumbel for tree 1 (Kendall's $\tau=0.6$), rotated Gumbel for tree 2 (Kendall's $\tau=0.4$), and rotated Clayton for tree 3 (Kendall's $\tau=0.2$). For $j=3$, the D-vine is truncated at tree 3, we set rotated Gumbel for tree 1 (Kendall's $\tau=0.7$), rotated Clayton for tree 2 (Kendall's $\tau=0.5$), and rotated Joe copula for tree 3 (Kendall's $\tau=0.3$).

We consider a candidate set of bivariate copulas that contains the most widely used copulas in practice, including the Independence, Gaussian, Frank, (rotated) Clayton, (rotated) Gumbel, and (rotated) Joe copulas. For each D-vine, we perform the sequential selection procedure under sample size $n=500$ and $1000$, and for each setting, we repeat the numerical experiment 500 times.

Table \ref{treebytree} reports the percentage of experiments where the sequential selection procedure correctly identifies the true truncation order of the D-vines and the percentage of experiments where the procedure correctly selects true bivariate copulas for each tree. As suggested by the result, the sequential selection procedure performs well in both D-vine truncation order selection and bivariate copula selection. In addition, the performance is improving with the increase of sample size $n$.

\begin{table}[h]
	\centering
	\caption{Performance of the tree-by-tree sequential selection procedure for three different D-vines.}
	\begin{tabular}{ccccc}\hline\hline
		1st D-vine    & $\text{trunc.\ at tree 2}$   & $\text{tree 1}$  & $\text{tree 2}$  &  \\\hline
		$n=500$ & 0.998 & 0.916 & 0.934 &\\
		$n=1000$ & 0.994 & 0.982 & 0.970 & \\\hline
		2nd D-vine    & $\text{trunc.\ at tree 3}$   & $\text{tree 1}$  & $\text{tree 2}$  & $\text{tree 3}$   \\\hline
		$n=500$ & 0.990 & 0.924 & 0.908 & 0.990\\
		$n=1000$ & 0.996 & 0.984 & 0.994 & 1 \\\hline
		3rd D-vine    & $\text{trunc.\ at tree 3}$   & $\text{tree 1}$  & $\text{tree 2}$  & $\text{tree 3}$  \\\hline
		$n=500$ & 0.988 & 0.966 & 1 & 0.818 \\
		$n=1000$ & 1 & 0.996 & 1 & 0.906 \\\hline\hline
	\end{tabular}
	\label{treebytree}
\end{table}

\section{Dataset for Empirical Analysis}\label{sec:moreresults}
Table \ref{tab:count} summarizes the empirical frequency of insurance claim count by year for each peril type. The distribution of claim count of each peril is stable over time. Water and fire are the most common causes of losses, which explains the reason that the fund combines all other perils into one category. On average, the odds of having zero claims per year due to water, fire, and other perils are 83.81\%, 86.20\%, and 87.95\%, respectively.

\begin{table}[htbp]
  \centering
  \caption{Empirical frequency of claim count by peril and by year.}
\begin{tabular}{crrrrr}
\hline\hline
     \multicolumn{6}{c}{Peril: Water}           \\
        Count   &       2006 &       2007 &       2008 &       2009 &       2010 \\
\cline{2-6}
         0 &        863 &        837 &        866 &        867 &        837 \\
         1 &        111 &        116 &        100 &         99 &        128 \\
         2 &         17 &         35 &         27 &         30 &         23 \\
         3 &         14 &          9 &          6 &          5 &         10 \\
         4 &          0 &          7 &          4 &          3 &          5 \\
         5 &          3 &          3 &          5 &          3 &          7 \\
       $> 5$ &         11 &         12 &         11 &         12 &          9 \\
\hline
     \multicolumn{6}{c}{Peril: Fire}           \\
       Count    &       2006 &       2007 &       2008 &       2009 &       2010 \\
\cline{2-6}
         0 &        871 &        879 &        890 &        896 &        856 \\
         1 &        111 &        102 &         99 &         95 &        109 \\
         2 &         22 &         24 &         26 &         18 &         33 \\
         3 &          8 &         11 &          4 &          4 &         13 \\
         4 &          3 &          3 &          0 &          5 &          7 \\
         5 &          2 &          0 &          0 &          0 &          0 \\
       $>5$ &          2 &          0 &          0 &          1 &          1 \\
\hline
     \multicolumn{6}{c}{Peril: Other}           \\
        Count   &       2006 &       2007 &       2008 &       2009 &       2010 \\
\cline{2-6}
         0 &        937 &        879 &        904 &        929 &        832 \\
         1 &         63 &        109 &        100 &         70 &        143 \\
         2 &         12 &         22 &          6 &         13 &         26 \\
         3 &          2 &          2 &          3 &          3 &          6 \\
         4 &          0 &          3 &          1 &          0 &          3 \\
         5 &          0 &          0 &          1 &          0 &          1 \\
       $>5$ &          5 &          4 &          4 &          4 &          8 \\
\hline\hline
\end{tabular}
  \label{tab:count}%
\end{table}%

Table \ref{tab:covariates} presents the descriptive statistics of exogenous predictors that the fund uses for underwriting and ratemaking. There are two categorical variables, entity type and alarm credit. Entity type is time constant, indicating whether an insured buildings belongs to a city, county, school, town, village, or a miscellaneous entity such as fire stations. Alarm credit is time-varying, reflecting the discount in premium received by a policyholder based on the features of the fire alarm system in the building. Available levels of discounts are 0\%, 5\%, 10\%, and 15\%. In addition, there is a time-varying continuous predictor, the amount of coverage (in million dollars), which measures the risk exposure of the policyholder. Due to the skewness, we use the coverage amount in log scale for our analysis.

\begin{table}[htbp]
  \centering
  \caption{Descriptive statistics of exogenous predictors$^\dag$.}
\begin{tabular}{lrrrrr}
\hline\hline
           &       2006 &       2007 &       2008 &       2009 &       2010 \\
\cline{2-6}
Entity Type &      &      &       &       &    \\
      City &    14.60\% &    14.60\% &    14.60\% &    14.60\% &    14.60\% \\
    County &     6.10\% &     6.10\% &     6.10\% &     6.10\% &     6.10\% \\
    School &    29.10\% &    29.10\% &    29.10\% &    29.10\% &    29.10\% \\
      Town &    16.40\% &    16.40\% &    16.40\% &    16.40\% &    16.40\% \\
   Village &    23.10\% &    23.10\% &    23.10\% &    23.10\% &    23.10\% \\
      Misc &    10.70\% &    10.70\% &    10.70\% &    10.70\% &    10.70\% \\
   \hline
Alarm Credit &      &      &       &       &    \\
      AC00 &    55.00\% &    52.40\% &    48.20\% &    39.40\% &    29.80\% \\
      AC05 &     2.50\% &     2.60\% &     3.40\% &     5.40\% &     7.40\% \\
      AC10 &     4.50\% &     5.10\% &     5.00\% &     6.70\% &     8.40\% \\
      AC15 &    38.10\% &    39.90\% &    43.40\% &    48.60\% &    54.40\% \\
   \hline
  Coverage &      34.98 &      37.73 &      39.98 &      42.62 &      43.03 \\
           &      (95.35) &     (102.53) &     (108.14) &     (114.41) &     (118.02) \\
\hline\hline
\multicolumn{6}{l}{$^\dag$ Standard deviations are presented in parentheses.}
\end{tabular}
  \label{tab:covariates}%
\end{table}%

\begin{figure}[htp]
  \begin{center}
   \includegraphics[width=0.6\textwidth,angle=270]{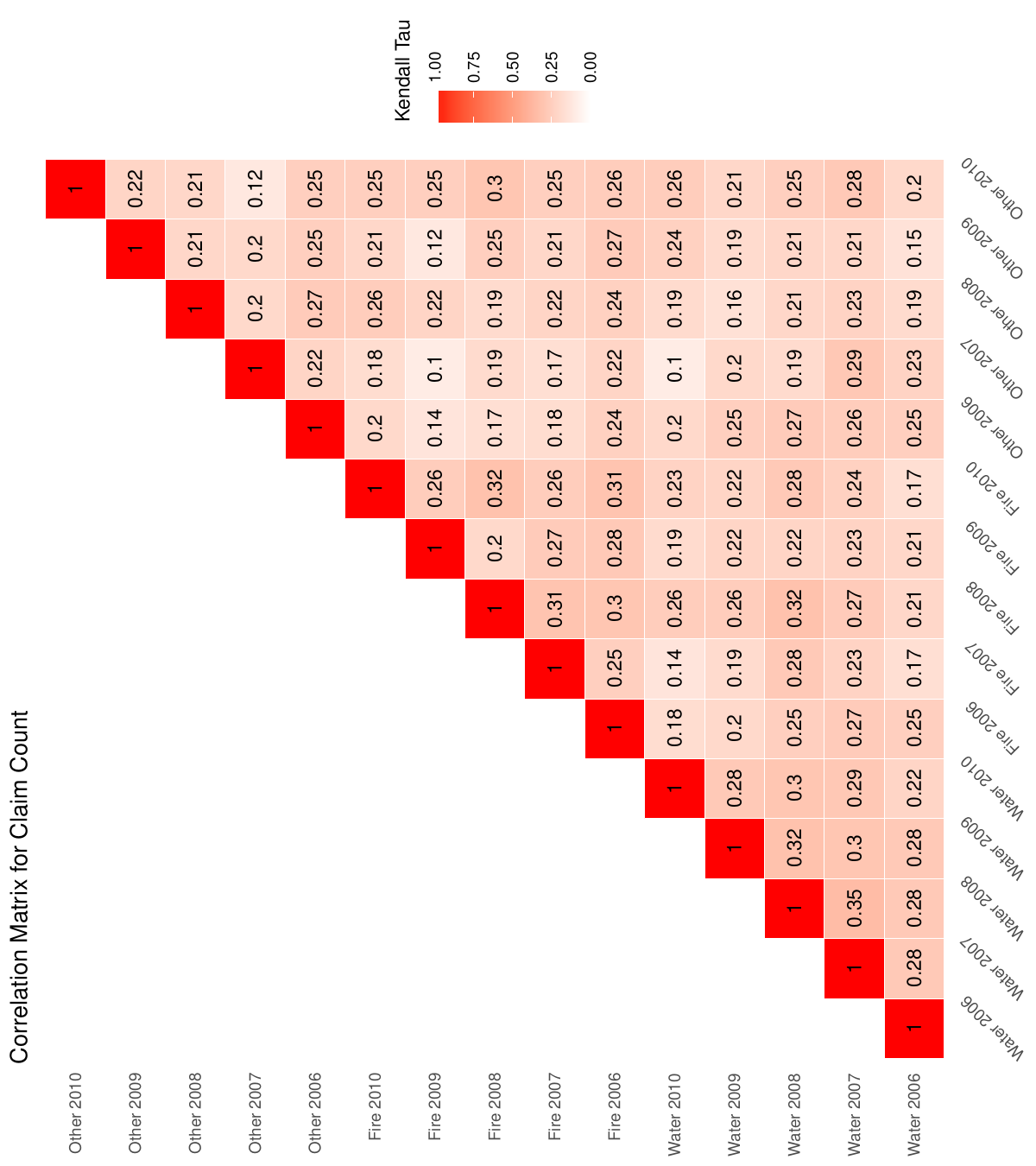}
  \end{center}
   \caption{Sample pairwise rank correlation~(i.e.\ Kendall's tau) matrix for the claim count.}
   \label{fig:corcount}
\end{figure}

To obtain intuitive knowledge of dependence among the bundled insurance risks, we visualize in Figure \ref{fig:corcount} the sample pairwise rank correlation~(i.e.\ Kendall's tau) for the multivariate longitudinal claim counts. Figure \ref{fig:corcount} suggests two sources of dependence. The first is the temporal dependence within each peril, which is due to unobserved peril-specific heterogeneity. This type of dependence is common for repeated measurements of the same risk over time, with measurements at closer time points being more correlated than measurements at more distant times. The second is the contemporaneous dependence among claim counts from different perils. This type of dependence is explained by unobserved policyholder-specific heterogeneity, i.e. measurements of multiple insurance risks realized on the same policyholder tend to be more alike than across policyholders.

{\color{black}
\section{Extension to Cross-policyholder Dependence}\label{suppsec:extension}
In this section, we provide an extension of the proposed D-vine based predictive model for multivariate longitudinal claim counts in Section \ref{subsec:riskmodel} of the main text to cover cross-policyholder dependence. 

Recall that the D-vine based predictive model (see equation \eqref{equ:cjoint} in the main text) assumes that, for each policyholder $i=1,2,\cdots,n$, we have 
\begin{align}\label{eq:original_model}
	F(\bm{y}_{it}|\bm{H}_{i,t-1}) = C^{J}\left(F\big(y_{it}^{(1)}|H_{i,t-1}^{(1)}\big),\ldots,F\big(y_{it}^{(J)}|H_{i,t-1}^{(J)}\big)\right),
\end{align}
where recall for the $i$th policyholder, $\bm{y}_{it}=(y_{it}^{(1)},\ldots,y_{it}^{(J)})$ denotes the claim counts of the $J$ perils at time $t$ and $\bm{H}_{it}=(H_{it}^{(1)},\ldots,H_{it}^{(J)})$ denotes the claim history up to time $t$. The conditional distribution $F\big(y_{it}^{(j)}|H_{i,t-1}^{(j)}\big)$ for $j=1,\cdots, J$ captures the temporal dependence within each peril and is specified by the univariate D-vine model in equation \eqref{equ:jtime} of the main text. A $J$-variate cross-sectional Gaussian copula $C^J$ with an unstructured $J\times J$ correlation matrix $\Sigma_J$ is employed to capture contemporaneous dependence among the $J$ perils within each policyholder.

Implicitly, the D-vine based predictive model assumes independence among different policyholders, which may not be realistic if the insurance policies are primarily designed to cover natural catastrophes (e.g.\ extreme temperature, windstorm, hail, flood), as policyholders located in the same spatial area may be subject to the same disaster and thus exhibit contemporaneous dependence. To address this, we propose to replace the $J$-variate copula $C^J$ with an $nJ$-variate copula $C^{*}$ that links and imposes contemporaneous dependence on $J$ perils among all $n$ policyholders.

In particular, the extended D-vine based predictive model takes the form
\begin{align}\label{eq:extended_model}
&F(\bm{y}_{1t}, \bm{y}_{2t}, \cdots, \bm{y}_{nt} |\bm{H}_{1,t-1},\cdots, \bm{H}_{n,t-1})\nonumber\\
 =& C^*\Big(F\big(y_{1t}^{(1)}|H_{1,t-1}^{(1)}\big),\ldots,F\big(y_{1t}^{(J)}|H_{1,t-1}^{(J)}\big),F\big(y_{2t}^{(1)}|H_{2,t-1}^{(1)}\big),\ldots,F\big(y_{2t}^{(J)}|H_{2,t-1}^{(J)}\big),\cdots\nonumber\\
 &\hspace{8mm}\cdots, F\big(y_{nt}^{(1)}|H_{n,t-1}^{(1)}\big),\ldots,F\big(y_{nt}^{(J)}|H_{n,t-1}^{(J)}\big)\Big),
\end{align}
where same as before, the conditional distribution $F\big(y_{it}^{(j)}|H_{i,t-1}^{(j)}\big)$ follows the univariate D-vine model in equation \eqref{equ:jtime} of the main text, while $C^{*}$ is an $nJ$-variate Gaussian copula with an $nJ\times nJ$ correlation matrix $\Sigma^*$ designed to capture the contemporaneous dependence of $J$ perils among all $n$ policyholders.

Note that suppose $\Sigma^*= I_{n} \otimes \Sigma^J$, where $I_{n}$ is an $n\times n$ identity matrix, $\Sigma_J$ is the $J\times J$ correlation matrix of the Gaussian copula $C^J$, and $\otimes$ is the Kronecker product, we have that \eqref{eq:extended_model} reduces to
\begin{align*}
	F(\bm{y}_{1t}, \bm{y}_{2t}, \cdots, \bm{y}_{nt} |\bm{H}_{1,t-1},\cdots, \bm{H}_{n,t-1})
	=\prod_{i=1}^{n} C^{J}\left(F\big(y_{it}^{(1)}|H_{i,t-1}^{(1)}\big),\ldots,F\big(y_{it}^{(J)}|H_{i,t-1}^{(J)}\big)\right),
\end{align*}
which is the D-vine based predictive model in the main text that assumes contemporaneous independence among different policyholders. 


\subsection{A Spatial Gaussian Copula $C^*$}
In this subsection, we design the correlation matrix $\Sigma^*$ of the copula $C^*$ based on a spatial Gaussian process, which allows flexible contemporaneous dependence among different policyholders.


In particular, we consider the following spatial Gaussian process:
\begin{align}\label{eq:Gprocess}
	z_{i}^{(j)}= \alpha_j s_{i}^{(j)} + \rho_j v_{i} + \varepsilon_{i,j}, ~\text{ for } i=1,\cdots,n;~ j=1,\cdots,J.
\end{align}
Here, $v_{i}$ and $\varepsilon_{i,j}$ are i.i.d.\ $N(0,1)$ across all indices, where $v_{i}$ is used to induce the policyholder effect and $\varepsilon_{i,j}$ is the idiosyncratic noise. Denote $\textbf{S}^{(j)}=\{s_{i}^{(j)}\}_{i=1}^n$, which is used to induce contemporaneous dependence among the $j$th peril of different policyholders. Specifically, we assume that $\textbf{S}^{(j)}$ is a unit-variance Gaussian process with an exponential covariance function (a special case of the Mat\'ern class) such that
\begin{align}\label{eq:spatial_cor}
	\mathrm{Cov}(s_{i_1}^{(j)}, s_{i_2}^{(j)})=\exp(-d_{i_1i_2}/\psi_j),
\end{align} 
where $d_{i_1i_2}$ is the spatial distance between policyholders $i_1$ and $i_2$, and can be calculated, for instance, based on the centroids. The parameter $\psi_j>0$ governs the strength of the spatial dependence. We assume that $\textbf{S}^{(j_1)}$ and $\textbf{S}^{(j_2)}$ are independent for $j_1\neq j_2$.

We set the $nJ$-variate spatial Gaussian copula $C^*$ as the one implied by the spatial Gaussian process $(z_{1}^{(1)},\cdots, z_{1}^{(J)}, z_{2}^{(1)},\cdots, z_{2}^{(J)}, \cdots \cdots, z_{n}^{(1)},\cdots, z_{n}^{(J)})$ in \eqref{eq:Gprocess}. Therefore, the correlation matrix $\Sigma^*$ of $C^*$ takes the form
\begin{align}\label{eq:spatial_cop}
	\Sigma^*_{(i_1,j_1),(i_2,j_2)}= \mathrm{Cor}(z_{i_1}^{(j_1)}, z_{i_2}^{(j_2)})=
	\begin{cases}
		1 & \text{ if } i_1=i_2, j_1=j_2,\\
		\dfrac{\rho_{j_1}\rho_{j_2}}{\sqrt{1+\alpha_{j_1}^2+\rho_{j_1}^2}\sqrt{1+\alpha_{j_2}^2+\rho_{j_2}^2}} & \text{ if } i_1=i_2, j_1\neq j_2,\\
		\dfrac{\alpha_j^2}{{1+\alpha_j^2+\rho_j^2}}\cdot \exp\big(-{d_{i_1i_2}}/{\psi_j}\big) &\text{ if } i_1\neq  i_2, j_1=j_2=j,\\
		0 & \text{ if } i_1\neq i_2, j_1\neq j_2,
	\end{cases}
\end{align}
where for notational simplicity, $\Sigma^*_{(i_1,j_1),(i_2,j_2)}$ denotes the $[(i_1-1)J+j_1, (i_2-1)J+j_2]$ entry of $\Sigma^*$. The model parameter of $\Sigma^*$ is thus $\boldsymbol{\rho}=\{(\alpha_j,\rho_j,\psi_j)\}_{j=1}^J.$ 
\medskip

\noindent\textbf{Remark S.1}: More flexible specification of $\Sigma^*$, and thus the spatial Gaussian copula $C^*$, can be achieved by modifying the spatial covariance structure of $\textbf{S}^{(j)}$ in \eqref{eq:spatial_cor}. For example, we can use the Mat\'ern covariance function or a spatial covariance function with a nugget effect if the policyholders can be further grouped into spatial clusters such as counties.
\medskip

\noindent\textbf{Remark S.2}: If $\alpha_j\equiv 0$ for all $j=1,\cdots, J$, there is no spatial dependence and $\Sigma^*$ reduces to 
\begin{align*}
	\Sigma^*= I_n \otimes \Sigma^{J*},
\end{align*}
where $\Sigma^{J*}$ is a factor-structured $J\times J$ correlation matrix that takes the form 
\begin{align*}
	\Sigma^{J*}_{j_1,j_2}=\frac{\rho_{j_1}\rho_{j_2}}{\sqrt{1+\rho_{j_1}^2}\sqrt{1+\rho_{j_2}^2}} \text{ for } j_1\neq j_2.
\end{align*}
Therefore, there is no contemporaneous dependence among policyholders and the extended model reduces to the D-vine based predictive model in the main text, where the Gaussian copula $C^J$ takes a factor-structured correlation matrix $\Sigma^{J*}$.

\subsection{Parameter Estimation of $C^*$}
Same as the D-vine based predictive model in Section \ref{subsec:riskmodel} of the main text, the extended model in \eqref{eq:extended_model} has three types of model parameters: the parameter $\bfbeta=(\bfbeta_1,\cdots,\bfbeta_J)$ of the marginal count regressions, the parameter $\bfzeta=(\bfzeta_1,\cdots,\bfzeta_J)$ of the D-vines, and the parameter $\boldsymbol{\rho}=\{(\alpha_j,\rho_j,\psi_j)\}_{j=1}^J$ of the spatial Gaussian copula $C^*$. We collect the model parameters as $\bftheta=(\bfbeta,\bfzeta,\bfrho)$.

Note that the two models share the same marginal count regressions and the same D-vines for temporal dependence, and the only difference is the cross-sectional copula $C^*$ for contemporaneous dependence. Therefore, it is easy to see that the three-stage MLE proposed in Section \ref{subsec:est} of the main text can still be used to estimate $\bfbeta$ of the marginal count regressions (in the first stage) and $\bfzeta$ of the D-vines (in the second stage) for the extended model.

However, to estimate $\bfrho$ of the spatial Gaussian copula $C^*$ for the extended model, the third stage estimator requires modification. In particular, due to the contemporaneous dependence among policyholders induced by $C^*$, the full log-likelihood of the extended model cannot be decomposed into the sum of log-likelihood of $n$ policyholders. Instead, given a portfolio of $n$ policyholders observed for $T$ periods $\{(\bm{y}_{i1},\ldots,\bm{y}_{iT})\}_{i=1}^n$, the full log-likelihood function can {only} be written as
\begin{align} \label{equ:loglik1}
	L(\bm{\theta}) = L(\bfbeta,\bfzeta,\bfrho) = \sum_{t=1}^{T}\log f(\bm{y}_{1t}, \bm{y}_{2t},\cdots, \bm{y}_{nt} |\bm{H}_{1,t-1},\cdots, \bm{H}_{n,t-1}),
\end{align}
where $\log f(\bm{y}_{1t}, \bm{y}_{2t},\cdots, \bm{y}_{nt} |\bm{H}_{1,t-1},\cdots, \bm{H}_{n,t-1})$ is the conditional log-likelihood of $J$ perils of all $n$ policyholders at time $t$ and {cannot} be further decomposed as $\sum_{i=1}^{n} \log f(\bm{y}_{it} |\bm{H}_{i,t-1})$ due to contemporaneous dependence among policyholders. 

The probability mass function~(pmf) $f(\bm{y}_{1t}, \bm{y}_{2t},\cdots, \bm{y}_{nt} |\bm{H}_{1,t-1},\cdots, \bm{H}_{n,t-1})$ in \eqref{equ:loglik1} needs to be computed based on the cumulative distribution function $F(\bm{y}_{1t}, \bm{y}_{2t},\cdots, \bm{y}_{nt} |\bm{H}_{1,t-1},\cdots, \bm{H}_{n,t-1})$ in \eqref{eq:extended_model}. This computation requires $2^{nJ}$ operations as the pmf is $nJ$-variate, which is computationally infeasible for moderate $nJ$. (Recall that computing the $J$-variate pmf $f(\bm{y}_{it} |\bm{H}_{i,t-1})$ requires $2^J$ operations, see \eqref{equ:jtyped} in the main text.) Recall that the three-stage MLE in Section \ref{subsec:est} of the main text estimates $\bfrho$ via
\begin{align*}
	\widehat{\bfrho}=\argmax L(\widehat{\bfbeta},\widehat{\bfzeta},\bfrho),
\end{align*}
where $\widehat{\bfbeta}$ and $\widehat{\bfzeta}$ are model parameters estimated in the first and second stage. Clearly, this estimator $\widehat{\bfrho}$ is infeasible due to the computational cost of the full log-likelihood \eqref{equ:loglik1}.

\textbf{A pairwise likelihood based estimator}: Instead, we consider the composite likelihood method~\citep{Lindsay1988}. The central idea is to use a (computationally efficient) lower-dimensional likelihood function to approximate the (computationally infeasible) full likelihood function. We refer readers to the recent review of \cite{Varin2011} and the references therein for more details. One important special case is the pairwise likelihood, which is defined based on the bivariate likelihood for pairs of observations. Owing to its computational efficiency and attractive asymptotic properties, the pairwise likelihood method has received much attention in the recent literature of spatial statistics and is used for estimation of spatial models developed for geocoded datasets~\citep[e.g.][]{Bevilacqua2012, HuserDavison2014, zhao2019composite, zhao2021knowledge}.

In particular, the pairwise likelihood of the extended model can be written as
\begin{align} \label{equ:loglik2}
	L_p(\bm{\theta}) = L_p(\bfbeta,\bfzeta,\bfrho) = \sum_{t=1}^{T}\sum_{i_1=1}^n\bigg\{&\sum_{i_2>i_1}\sum_{j_1=1}^J\sum_{j_2=1}^J\log f(y_{i_1,t}^{(j_1)}, y_{i_2,t}^{(j_2)} |{H}_{i_1,t-1}^{(j_1)}, {H}_{i_2,t-1}^{(j_2)})+\nonumber\\
	& \sum_{j_1=1}^J\sum_{j_2>j_1}\log f(y_{i_1,t}^{(j_1)}, y_{i_1,t}^{(j_2)} |{H}_{i_1,t-1}^{(j_1)}, {H}_{i_1,t-1}^{(j_2)})\bigg\}.
\end{align}
For any $(i_1,j_1)$ and $(i_2,j_2)$, the pairwise likelihood can be computed via
\begin{align*}
	&f(y_{i_1,t}^{(j_1)}, y_{i_2,t}^{(j_2)} |{H}_{i_1,t-1}^{(j_1)}, {H}_{i_2,t-1}^{(j_2)})\nonumber\\=& \sum_{k_1=0}^{1}\sum_{k_2=0}^{1}(-1)^{k_1+k_2}C_{(i_1,j_1),(i_2,j_2)}\left(F\left(y_{i_1,t}^{(j_1)}-k_1|H_{i_1,t-1}^{(j_1)}\right),F\left(y_{i_2,t}^{(j_2)}-k_2|H_{i_2,t-1}^{(j_2)}\right)\right),
\end{align*}
with the convention $F(y|H_{i,t-1}^{(j)})=0$ for $y<0$, where $C_{(i_1,j_1),(i_2,j_2)}$ is a bivariate Gaussian copula with correlation $\Sigma^*_{(i_1,j_1),(i_2,j_2)}$ as specified in \eqref{eq:spatial_cop}.

Thanks to the use of pairwise likelihood, the likelihood function \eqref{equ:loglik2} can be efficiently computed. Therefore, for the extended model, we propose to estimate $\bfrho$ in the third stage via
\begin{align*}
	\widetilde{\bfrho}=\argmax L_p(\widehat{\bfbeta},\widehat{\bfzeta},\bfrho),
\end{align*}
where $\widehat{\bfbeta}$ and $\widehat{\bfzeta}$ are model parameters estimated in the first and second stage based on the three-stage MLE in Section \ref{subsec:est} of the main text.

It is known that a pairwise likelihood based estimator is consistent and asymptotically normal, though in general it is less efficient than the classical MLE due to mis-specification of the true likelihood. In particular, theoretical guarantees for $\widetilde{\bfrho}$ can be established using similar technical arguments as the ones for $\widehat{\bfrho}$ and therefore are omitted. 
}

\bibliographystyle{apalike}
\bibliography{ref_MixedVine}

\end{appendices}